\documentclass[twocolumn]{aastex631}

\usepackage{amsmath}

\received{8-Sep-2022}
\revised{21-Dec-2022}
\accepted{23-Jan-2023}

\submitjournal{ApJ}

\shorttitle{SN2020uem: (II) The Properties of The CSM}
\shortauthors{UNO, Nagao, \& Maeda et al.}

\graphicspath{{./}{figures/}}

\begin{document}

\title{SN 2020uem: A Possible Thermonuclear Explosion within A Dense Circumstellar Medium\\ 
(II) The Properties of The CSM from Polarimetry and Light Curve Modeling}

\correspondingauthor{Kohki Uno}
\email{k.uno@kusastro.kyoto-u.ac.jp}

\author[0000-0002-6765-8988]{Kohki Uno}
\affiliation{Department of Astronomy, Kyoto University, Kitashirakawa-Oiwake-cho, Sakyo-ku, Kyoto, 606-8502, Japan}

\author[0000-0002-3933-7861]{Takashi Nagao}
\affiliation{Department of Physics and Astronomy, University of Turku, FI-20014 Turku, Finland}

\author[0000-0003-2611-7269]{Keiichi Maeda}
\affiliation{Department of Astronomy, Kyoto University, Kitashirakawa-Oiwake-cho, Sakyo-ku, Kyoto, 606-8502, Japan}

\author[0000-0002-1132-1366]{Hanindyo Kuncarayakti}
\affiliation{Department of Physics and Astronomy, University of Turku, FI-20014 Turku, Finland}
\affiliation{Finnish Centre for Astronomy with ESO (FINCA), FI-20014 University of Turku, Finland}

\author[0000-0001-8253-6850]{Masaomi Tanaka}
\affiliation{Astronomical Institute, Tohoku University, Sendai 980-8578, Japan}
\affiliation{Division for the Establishment of Frontier Sciences, Organization for Advanced Studies, Tohoku University, Sendai 980-8577, Japan}

\author[0000-0001-6099-9539]{Koji S. Kawabata}
\affiliation{Hiroshima Astrophysical Science Center, Hiroshima University, Kagamiyama 1-3-1, Higashi-Hiroshima ,Hiroshima 739-8526, Japan}
\affiliation{Department of Physical Science, Hiroshima University, Kagamiyama 1-3-1, Higashi-Hiroshima 739-8526, Japan}

\author{Tatsuya Nakaoka}
\affiliation{Hiroshima Astrophysical Science Center, Hiroshima University, Kagamiyama 1-3-1, Higashi-Hiroshima ,Hiroshima 739-8526, Japan}
\affiliation{Department of Physical Science, Hiroshima University, Kagamiyama 1-3-1, Higashi-Hiroshima 739-8526, Japan}

\author[0000-0002-4540-4928]{Miho Kawabata}
\affiliation{Okayama Observatory, Kyoto University, 3037-5 Honjo, Kamogatacho, Asakuchi, Okayama 719-0232, Japan}

\author[0000-0001-9456-3709]{Masayuki Yamanaka}
\affiliation{Okayama Observatory, Kyoto University, 3037-5 Honjo, Kamogatacho, Asakuchi, Okayama 719-0232, Japan}

\author[0000-0003-4569-1098]{Kentaro Aoki} 
\affiliation{Subaru Telescope, National Astronomical Observatory of Japan, 650 North A’ohoku Place, Hilo, HI 96720, USA}

\author{Keisuke Isogai}
\affiliation{Okayama Observatory, Kyoto University, 3037-5 Honjo, Kamogatacho, Asakuchi, Okayama 719-0232, Japan}
\affiliation{Department of Multi-Disciplinary Sciences, Graduate School of Arts and Sciences, The University of Tokyo, 3-8-1 Komaba, Meguro, Tokyo 153-8902, Japan}

\author[0000-0001-5822-1672]{Mao Ogawa}
\affiliation{Department of Astronomy, Kyoto University, Kitashirakawa-Oiwake-cho, Sakyo-ku, Kyoto, 606-8502, Japan}

\author[0000-0001-8813-9338]{Akito Tajitsu} 
\affiliation{Okayama Branch Office, Subaru Telescope, National Astronomical Observatory of Japan, Kamogata, Asakuchi, Okayama, 719-0232, Japan}

\author[0000-0002-0643-7946]{Ryo Imazawa}
\affiliation{Department of Physics, Graduate School of Advanced Science and Engineering, Hiroshima University, Kagamiyama 1-3-1, Higashi-Hiroshima ,Hiroshima 739-8526, Japan}

\begin{abstract}

Type IIn/Ia-CSM supernovae (SNe IIn/Ia-CSM) are classified by their characteristic spectra, which exhibit narrow hydrogen emission lines originating from a strong interaction with a circumstellar medium (CSM) together with broad lines of intermediate-mass elements. We performed intensive follow-up observations of SN IIn/Ia-CSM 2020uem, including photometry, spectroscopy, and polarimetry. In this paper, we focus on the results of polarimetry. We performed imaging polarimetry at $66$ days and spectropolarimetry at $103$ days after the discovery. SN 2020uem shows a high continuum polarization of $1.0-1.5\%$ without wavelength dependence. Besides, the polarization degree and position angle keep roughly constant. These results suggest that SN 2020uem is powered by a strong interaction with a confined and aspherical CSM. We performed a simple polarization modeling, based on which we suggest that SN 2020uem has an equatorial-disk/torus CSM. Besides, we performed semi-analytic light-curve modeling and estimated the CSM mass. We revealed that the mass-loss rate in the final few hundred years immediately before the explosion of SN 2020uem is in the range of $0.01 - 0.05 {\rm ~M_{\odot}~yr^{-1}}$, and that the total CSM mass is $0.5-4 {\rm ~M_{\odot}}$. The CSM mass can be accommodated by not only a red supergiant (RSG) but a red giant (RG) or an asymptotic-giant-branch (AGB) star. As a possible progenitor scenario of SN 2020uem, we propose a white-dwarf binary system including an RG, RSG or AGB star, especially a merger scenario via common envelope evolution, i.e., the core-degenerate scenario or its variant.

\end{abstract}


\section{Introduction} \label{sec:1}

Supernovae (SNe) are one of the endpoints of stellar evolution. Based on the spectroscopic and photometric properties, SNe are classified into several subclasses \citep{1997ARA&A..35..309F}. Type Ia SNe (SNe Ia) show distinctive spectral evolution characterized by the absence of hydrogen and helium lines and by the presence of strong absorption features of intermediate-mass elements, e.g., silicon and sulfur. SNe Ia are triggered by a thermonuclear explosion of a white dwarf (WD) when its mass exceeds a threshold. The channel leading to the mass threshold remains unclear \citep[e.g.,][for a review]{Maeda2016IJMPD}.  One of the leading scenarios toward the explosion is the mass accretion onto the WD from a non-degenerate companion star \citep[single-degenerate (SD) scenario; e.g.,][]{Whelan1973ApJ,Nomoto1982ApJ}. Another popular scenario is the merger of two sub-Chandrasekhar WDs \citep[double-degenerate (DD) scenario; e.g.,][]{Webbink1984ApJ,Iben1984ApJS}.

In the recent decades, peculiar SNe Ia, which are referred to as Type Ia-CSM SNe (SNe Ia-CSM), have been discovered. SNe Ia-CSM are characterized by the narrow hydrogen emission lines similar to Type IIn SNe (SNe IIn), on the top of a continuum with broad features of Fe-peak elements similar to SNe Ia. Besides, SNe Ia-CSM keep high luminosity ($\gtrsim 10^{42-43}{\rm ~erg~s^{-1}}$) over $\sim 100$ days. These observational properties suggest that SNe Ia-CSM are powered by an interaction between the ejecta of a WD explosion and a dense circumstellar medium (CSM). However, in the classical picture of stellar evolution, WDs are terminal points of the low-mass stars ($\lesssim 8 {\rm ~M_{\odot}}$). Therefore, it is difficult to form such a dense CSM around a WD. The nature and origin of the CSM still remain unclear.

To explain the WD explosion in the dense CSM, some progenitor scenarios, e.g., a stellar merger via common envelope (CE) evolution \citep[core-degenerate (CD) scenario and its variant; e.g.,][]{Livio2003ApJ, Kashi2011MNRAS, Jerkstrand2020Science} and a thermonuclear explosion of a degenerate core of an asymptotic-giant-branch (AGB) star \citep[Type 1.5 SNe; e.g.,][]{Hamuy2003Nature}, have been proposed. However, the nature of the progenitor system, especially as to whether it involves a WD, has still been controversy \citep[e.g.,][]{Benetti2006ApJ}. The controversy is further added by a difficulty to differentiate `genuine' SNe Ia-CSM from SNe IIn, given their similar observational characteristics. Therefore, SNe Ia-CSM `candidates' are frequently termed SNe IIn/Ia-CSM; some suggest that at least a fraction of SNe IIn/Ia-CSM originate from core-collapse SNe of massive stars \citep[e.g.,][]{Inserra2014MNRAS}. 

The sample of SNe Ia-CSM and IIn/Ia-CSM is still limited, e.g., SNe Ia-CSM; SN 2002ic \citep{Hamuy2003Nature, Deng2004ApJ, Wang2004ApJ}, PTF11kx \citep{Dilday2012Science}, and SNe IIn/Ia-CSM; SN 2005gj \citep{Aldering2006ApJ, Prieto2007arxiv}, SN 2008J \citep{Taddia2012AA}, SN 2012ca \citep{Inserra2014MNRAS,Fox2015MNRAS,Inserra2016MNRAS}, and SN 2013dn \citep{Fox2015MNRAS}.
To constrain the progenitor systems of SNe IIn/Ia-CSM, deriving the CSM properties, e.g., the CSM mass, geometry, and mass-loss history, is important. In particular, the CSM geometry potentially provides us with key implications for the mechanisms of the CSM formation. Polarization gives us a direct clue to reveal the CSM geometry. Indeed, normal core-collapse and thermonuclear SNe, which are overall spherical explosions without dense CSMs, generally show low polarization degrees of $\lesssim 0.2\%$ \citep{Yang2020ApJ} in their continuum spectra, with a few exceptions \citep{Nagao2021MNRAS}. On the other hand, strongly-interacting SNe tend to exhibit high polarization degrees of $\gtrsim 1\%$, which is probably caused by aspherical CSMs \citep[e.g.,][]{Patat2011AA,Reilly2017MNRAS,Kumar2019MNRAS}. However, the polarization sample of SNe IIn/Ia-CSM has been very limited \citep[e.g.,][]{Wang2004ApJ}, and no detailed analysis has been performed. 

SN 2020uem/ATLAS20bbsz is an SNe IIn/Ia-CSM, which was discovered on September 22.602 2020 UT (MJD 59114.602) by the Asteroid Terrestrial-impact Last Alert System \citep[ATLAS;][]{TonryPASP2018, Tonry2020TNS} project. The coordinate is $\rm{RA(J2000.0)} = 08^{\rm h}24^{\rm m}23^{\rm s}.85$ and ${\rm Dec(J2000.0)} = -03^{\circ}29^{\prime}19^{\prime\prime}.1$. SN 2020uem is located at $z = 0.041$ ($d_{\rm L} = 173.3 \pm 5.7 {\rm ~ Mpc}$). We have performed intensive follow-up observations of SN 2020uem. In addition to optical/near-infrared (NIR) photometry and spectroscopy, we have also obtained polarization data. Our dataset contains both imaging polarimetry and spectropolarimetry. The results of optical/NIR photometry and spectroscopy have been presented by Uno et al. (Paper I). In this paper, we focus on the results of polarimetry. Besides, taking the results of polarimetry into account, we performed light curve modeling with a semi-analytical model \citep[based on][]{Moriya2013MNRAS_b,Nagao2020MNRAS}.

This paper is structured as follows. In Section \ref{sec:2}, we summarize our reduction processes of polarimetry. In Section \ref{sec:3}, we show the results of spectropolarimetry, followed by discussion on the time evolution of polarization. In Section \ref{sec:4}, with a polarization modeling based on the results of  \citet{Hoflich1991AA}, we constrain the properties of the CSM geometry and suggest that a torus CSM is feasible for SN 2020uem. In Section \ref{sec:5}, assuming the torus CSM, we study the light curve of SN 2020uem with a semi-analytical model and constrain the CSM mass. This paper is closed in Section \ref{sec:6} with discussion and conclusions.

\section{Observations and Data Reduction} \label{sec:2}

\subsection{Spectropolarimetry with FOCAS} \label{sec:2.1}

The spectropolarimetry of SN 2020uem with the Faint Object Camera and Spectrograph \citep[FOCAS;][]{2002PASJ...54..819K} on the Subaru telescope was performed on January 3.20 2021 UT (MJD 59217.20) and 4.15 2021 UT (MJD 59218.15). We use the B300 grating with $0.8^{\prime\prime}$ center-slit and no order-sorting filter. Under this configuration, the wavelength coverage is $3650-8300$ {\text \AA} with a spectral resolution of $R = \lambda/\Delta \lambda \sim 500$. Given that the FOCAS is attached to the Cassegrain focus and the slit is placed in a symmetric way, the instrumental polarization is reduced to be minimal. While the second-order scattering light may (slightly) contaminate the signal above $\sim 7000$ {\text \AA}, we note that any of our conclusions would not rely on the data in this particular wavelength range. 

FOCAS has a Wollaston prism and a rotating half-wave plate (HWP). The Wollaston prism splits the incident ray into two beams with orthogonal polarization directions; ordinary and extraordinary beams. Our spectropolarimetric data with FOCAS are composed of four frames for one set, corresponding to the HWP rotation angles of $0^{\circ}$, $22.5^{\circ}$, $45^{\circ}$, and $67.5^{\circ}$. The exposure time for each frame was 450 seconds, i.e., the total exposure time for one set was 1800 seconds. We obtained four sets of spectropolarimetry. In addition,, we obtained the following two standard stars for calibrations: G191B2B and HD24553. We listed the summary of the spectropolarimetric observation in Table \ref{tab:spectropolarimerty}. 

We reduced the data with IRAF\footnote{IRAF is distributed by the National Optical Astronomy Observatory, which is operated by the Association of Universities for Research in Astronomy (AURA) under a cooperative agreement with the National Science Foundation.} using the standard procedure for spectropolarimetry \citep[for detail, see ][]{Patat2017SNhandbook}. We extracted the ordinary and extraordinary spectra from a single CCD image and calibrated their wavelength using the arc lamp (Th and Ar) data. For the flux calibration, we used G191B2B. Finally, we obtained the following data: $I_{0}$, $I_{45}$, $I_{90}$ and $I_{135}$, where $I_{\phi}$ is the flux where $\phi$ corresponds to twice the HWP angle. We defined the Stokes parameters $Q$, $U$ and $P$ and position angle $\theta$ as follows: $Q = (I_{0} - I_{90}) / (I_{0} + I_{90}) =  (I_{0} - I_{90}) / I =  P\cos 2\theta$, $U = (I_{45} - I_{135}) / (I_{45} + I_{135}) = (I_{45} - I_{135}) / I = P\sin 2\theta$, $P = \sqrt{Q^{2} + U^{2}}$, and $\theta = 1/2 \arctan\left( U/Q \right)$, where $I = I_{0} + I_{90} = I_{45} + I_{135}$ is the total flux. Note that we estimated the Stokes parameters by combining the data obtained in two consecutive nights, assuming that the variation of the parameters in one day is negligible because the characteristic timescale must be much longer in the late phase.

As the calibration of the Stokes parameters, some corrections were performed. First, using the non-polarization standard star (G191B2B), we calibrated the instrumental polarization of FOCAS, which is estimated to be $Q\sim 0.1-0.2\%$ and $U\sim 0.1-0.2\%$. Besides, we also calibrated the offset of the position angle from the reference axis on the celestial plane using the strongly-polarized standard star (HD24553), whose position angle is $135.1^{\circ} \pm 0.2^{\circ}$ \citep{Wolff1996AJ}. Then, we calibrated the wavelength dependence of the position angle using the fully-polarized flat lamp data. For the polarization bias correction, we use the standard method described in \citet{Wang1997ApJ} as follows:
\begin{equation}
    P_{\rm true} = P_{\rm obs} - \sigma^{2} / P_{\rm obs},
\end{equation}
where $P_{\rm true}$ and $P_{\rm obs}$ are polarization degrees after and before the bias correction, respectively, and $\sigma$ is the error of the polarization degree. Finally, we corrected the interstellar polarization (ISP, see section \ref{sec:3.1}).

\begin{deluxetable*}{cccccc}
\tablenum{1}
\tablecaption{Log of Spectropolarimerty of SN 2020uem with FOCAS\label{tab:spectropolarimerty}}
\tablewidth{0pt}
\tablehead{
\colhead{object} & \colhead{Date} & \colhead{MJD} & \colhead{Phase} & \colhead{Exposure Time} & \colhead{Type} \\
\nocolhead{} & \nocolhead{} & \nocolhead{} & \colhead{(day)} & \nocolhead{} & \colhead{}
}
\startdata
  SN 2020uem & 2021-01-03 & 59217.20 & 102.6 & (450 (sec) $\times$ 4 (angle)) $\times$ 2 set & Type IIn/Ia-CSM SNe \\
  SN 2020uem & 2021-01-04 & 59218.15 & 103.6 & (450 (sec) $\times$ 4 (angle)) $\times$ 2 set & Type IIn/Ia-CSM SNe \\
  G191B2B    & 2021-01-03 & 59217.20 & - & (60 (sec) $\times$ 4 (angle)) $\times$ 1 set &  unpolarized standard \\
  HD24553   & 2021-01-03 & 59217.20  & - & (1 (sec) $\times$ 4 (angle)) $\times$ 1 set &  polarized standard \\
\enddata
\tablecomments{The phase means days relative to the epoch of the discovery (MJD 59114.602).}
\end{deluxetable*}

\subsection{Imaging Polarimetry with Dipol-2} \label{sec:2.2}

The broad-band polarimetry data in the V and R bands have been acquired with the remotely controlled Dipol-2 polarimeter mounted on the Tohoku T60 telescope at Haleakala Observatory (Hawaii) during one night on November 27.05 2020 UT (MJD 59180.05) by Berdyugin (2020, private communication). The single exposure time was 60 seconds for each band, and the total integration time was 140 minutes. The mean values of the polarization and position angle, as computed by averaging 35 individual measurements, are the following: $(P,~\theta) = (2.44\%, ~77.6^{\circ})$ in the V band and $(P,~\theta) = (1.67\%, ~83.4^{\circ})$ in the R band. Description of the instrument characteristics, calibration and data reduction techniques are given in \citet{Piirola2014SPIE}, \citet{Kosenkov2017MNRAS}, and \citet{Piirola2020AA}.

\section{Polarimetry} \label{sec:3}

\subsection{Interstellar Polarization} \label{sec:3.1}

To estimate the ISP wavelength dependence, we use the Serkowski function \citep{Serkowski1975ApJ} described as follows:
\begin{equation}
    P(\lambda)=P_{\max } \exp \left[-K \ln ^{2}\left(\lambda_{\max } / \lambda\right)\right],
\end{equation}
where $\lambda_{\max}$ is the wavelength where the ISP reaches the maximum polarization degree ($P_{\max}$) and $K$ is given by \citet{Whittet1992ApJ} as follows: $K = 0.01 + 1.66\lambda_{\max} ~ ({\rm \mu m})$. For the ISP estimation, we adopt the polarization degrees at several prominent narrow emission lines, i.e., $\rm H \alpha$, $\rm H \beta$, and \ion{He}{1} $\lambda5876$, based on the FOCAS data. We see that $P(\lambda)$ takes the minimum value at the peak of each line. The position angles are consistent at the peaks of different lines, being $60-80^{\circ}$. We assume that the peaks of the narrow lines are not polarized by the SN ejecta and the CSM interaction since the narrow lines are formed by the recombination in the unshocked CSM. Therefore, the emission-line peaks are good tracers of the ISP.

\begin{figure*}
\epsscale{1.17}
\plotone{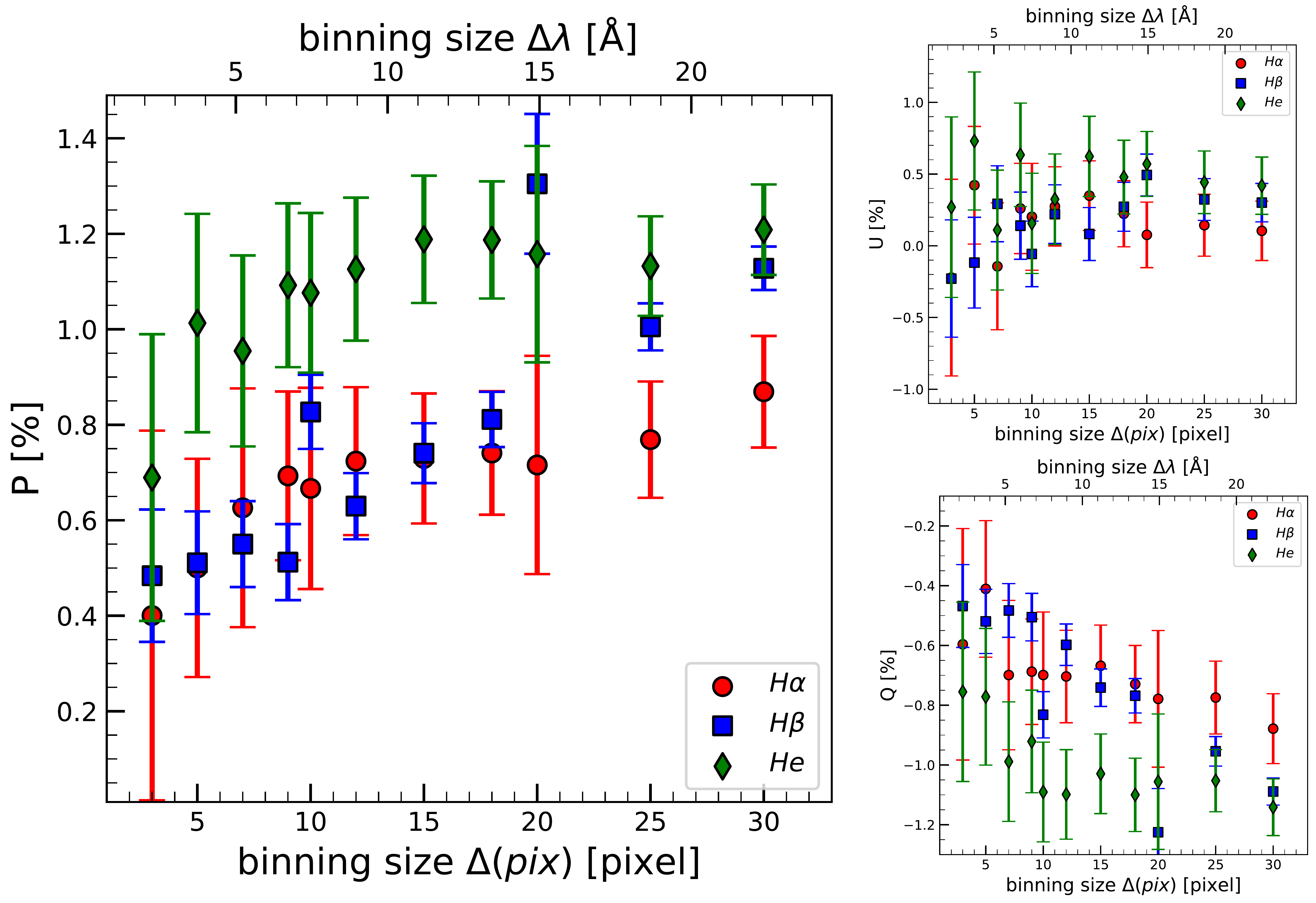}
\caption{The polarization degrees of $\rm H\alpha$, $\rm H\beta$, and \ion{He}{1} $\lambda5876$ as a function of the binning size. We sampled the Stokes parameters at the peaks of the three narrow emission lines. The error bars show the photon shot noise per bin.}
\label{fig:1}
\end{figure*}

In Figure \ref{fig:1}, in order to determine the optimal binning size to estimate the ISP, we plot the Stokes parameters at the peaks of the three narrow emission lines as a function of the binning size. Assuming that the line center is fully unpolarized while the line wing is a mixture of the unpolarized lines and the polarized component, we can correctly extract the unpolarized signal alone using the small binning size but with a large error due to the photon statistics, i.e., low signal-to-noise ratio (S/N). As the binning size is increased, the polarization degrees of the lines are still not much affected by the wing and thus the unpolarized level can be picked up, and the noise decreases. However, once a threshold in the binning size is reached, the polarized SN component starts contaminating to the unpolarized ISP component, and thus the polarization degree changes as a function of binning size while the S/N keeps increasing. Indeed, Figure \ref{fig:1} shows that the Stokes parameters is gradually varied for $\Delta \lambda \gtrsim 5$ {\text \AA}, and the parameters keep roughly constant at a binning size larger than $\sim 10$ {\text \AA}. Therefore, we adopt the 5-pixel-binned ($\approx 4$ {\text \AA}) spectrum to estimate the ISP.
  
We fit the Serkowski function to the ISP component at the wavelengths of the three lines, and then the best-fit parameters are obtained as follows: $\lambda_{\max} = 5235$ {\text \AA}, $P_{\max} = 0.86\%$, and $K = 0.10$. The Na D lines originated both from the host and the Milky Way (MW) are not detected, indicating that the host extinction is not significant. The polarization degrees of filed stars toward the direction of SN 2020uem shows a range of values, $\sim 0.1-1\%$ \citep{Heiles2000AJ}, covering the ISP value estimated for SN 2020uem. Thus, it is possible that the ISP is dominated by the MW ISP. In any case, the origin of the ISP being the MW or the host would not affect our main conclusions.

\subsection{Stokes Parameters of SN 2020uem} \label{sec:3.2}

\begin{figure*}
\gridline{\fig{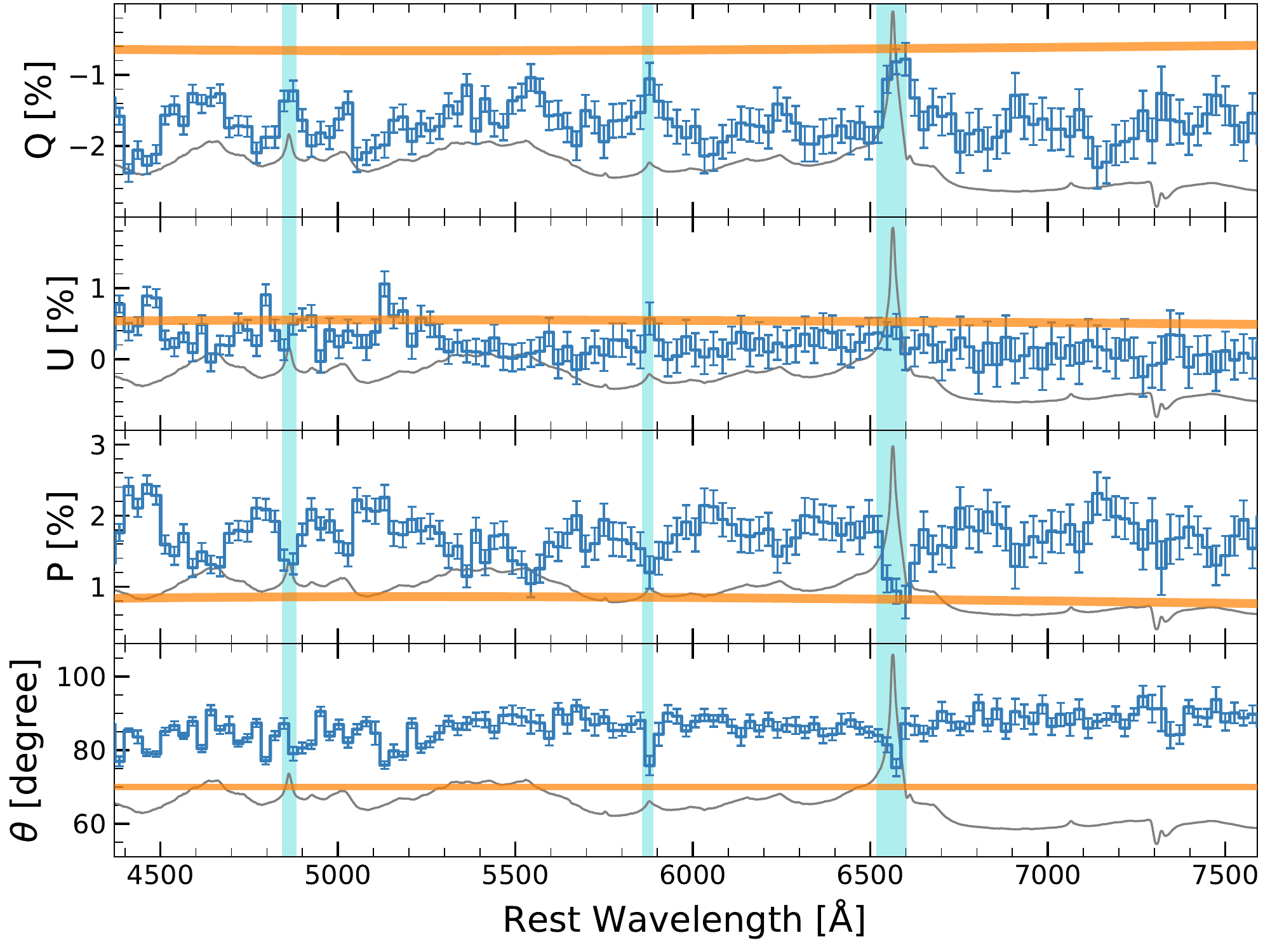}{0.45\textwidth}{non-ISP-subtracted data}
          \fig{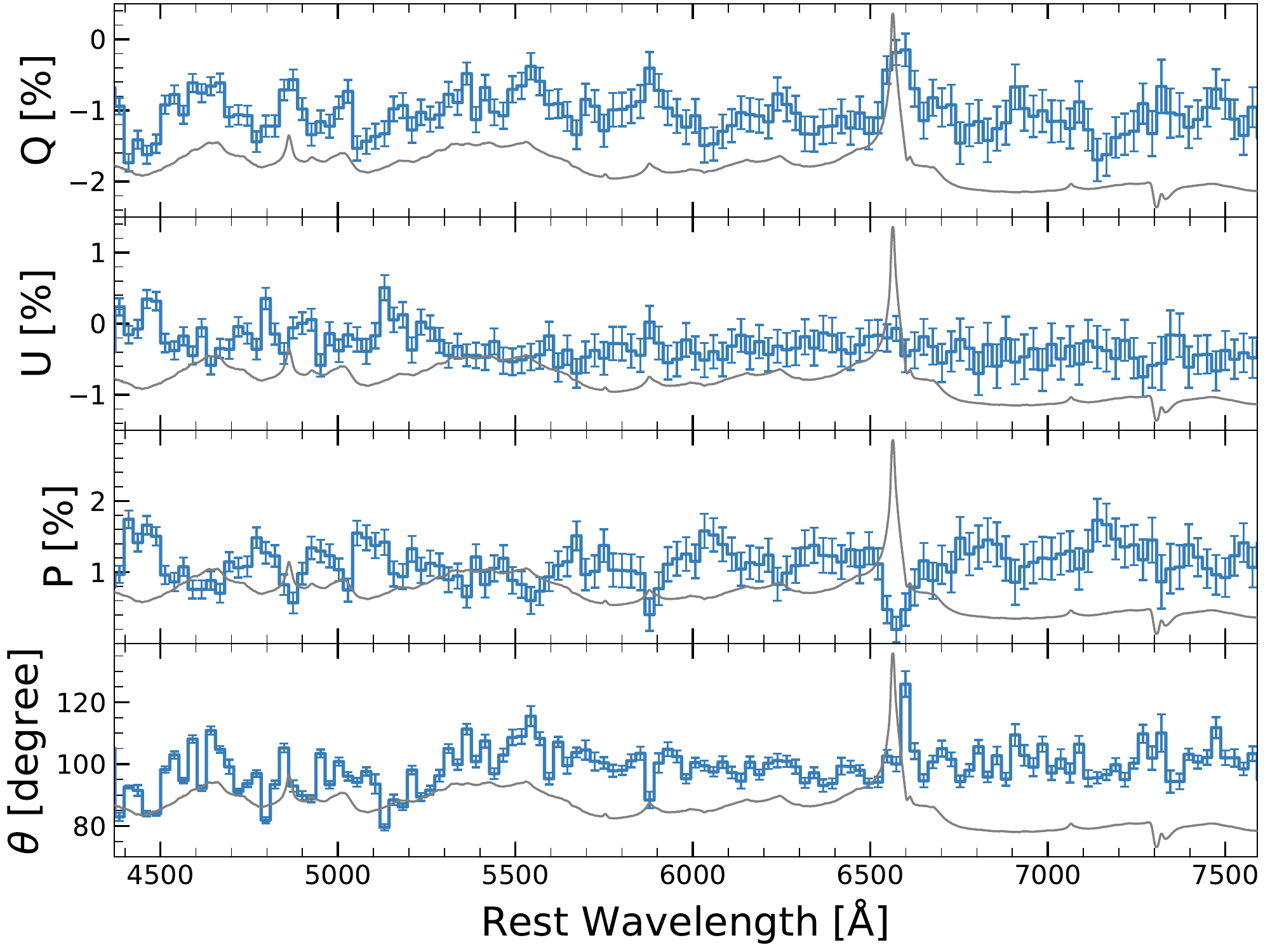}{0.45\textwidth}{ISP-subtracted data}
          }
\caption{Left Panel: the non-ISP-subtracted spectrum of SN 2020uem at $+103$ day from the discovery. Right Panel: the ISP-subtracted spectrum. From top to bottom panel, the Stokes parameters $Q$ and $U$, the total polarization $P$, and the position angle $\theta$ are plotted. The spectra shown by blue are binned by 20 pixels. In our configuration of FOCAS, the wavelength resolution per pixel is $\sim 1.34$ {\text \AA} for the unbinned spectrum, and then the resolution for the 20-pixel-binned spectrum corresponds to $\sim 26.8$ {\text \AA}. The error bars show the photon-shot noise per bin. The orange lines in the left panels show the estimated ISP by the Serkowski function. In the left-bottom panel, the orange line represents the assumed ISP angle ($\theta = 70^{\circ}$). We identify the light-blue shaded wavelengths, i.e., $\rm H \alpha$, $\rm H \beta$, and \ion{He}{1} $\lambda5876$, as the ISP-dominated components, and estimated the ISP by fitting those data with the Serkowski function. The grey spectrum in the background of each plot is the unbinned spectrum of SN 2020uem.}
\label{fig:2}
\end{figure*}

Figure \ref{fig:2} shows the Stokes parameters and position angle of SN 2020uem at $+103$ day after the discovery. SN 2020uem shows a strong-polarized continuum of $P \sim 1.0-1.5\%$. We note that SN 2002ic, as the prototypical SN Ia-CSM, showed a high polarization degree of $\sim 0.8\%$ \citep{Wang2004ApJ}, and SN 2020uem shows an even stronger polarization. The result here suggests an asymmetric geometry for the CSM around SN 2020uem, which is likely shared by SNe IIn/Ia-CSM as a common property. Besides, the Stokes parameters and the position angle, after subtracting the ISP component, do not show wavelength dependence in the continuum component. The absence of wavelength dependence implies that the continuum is dominated by a physical process independent of wavelength, i.e., electron scattering, which probably traces a dense and ionized CSM. Therefore, we conclude that SN 2020uem has an aspherical and dense CSM, e.g., an equatorial-disk/torus or clumpy CSM. A more quantitative discussion of the CSM structure is provided in Section \ref{sec:4} and \ref{sec:5}.

\begin{figure}
\epsscale{1.17}
\plotone{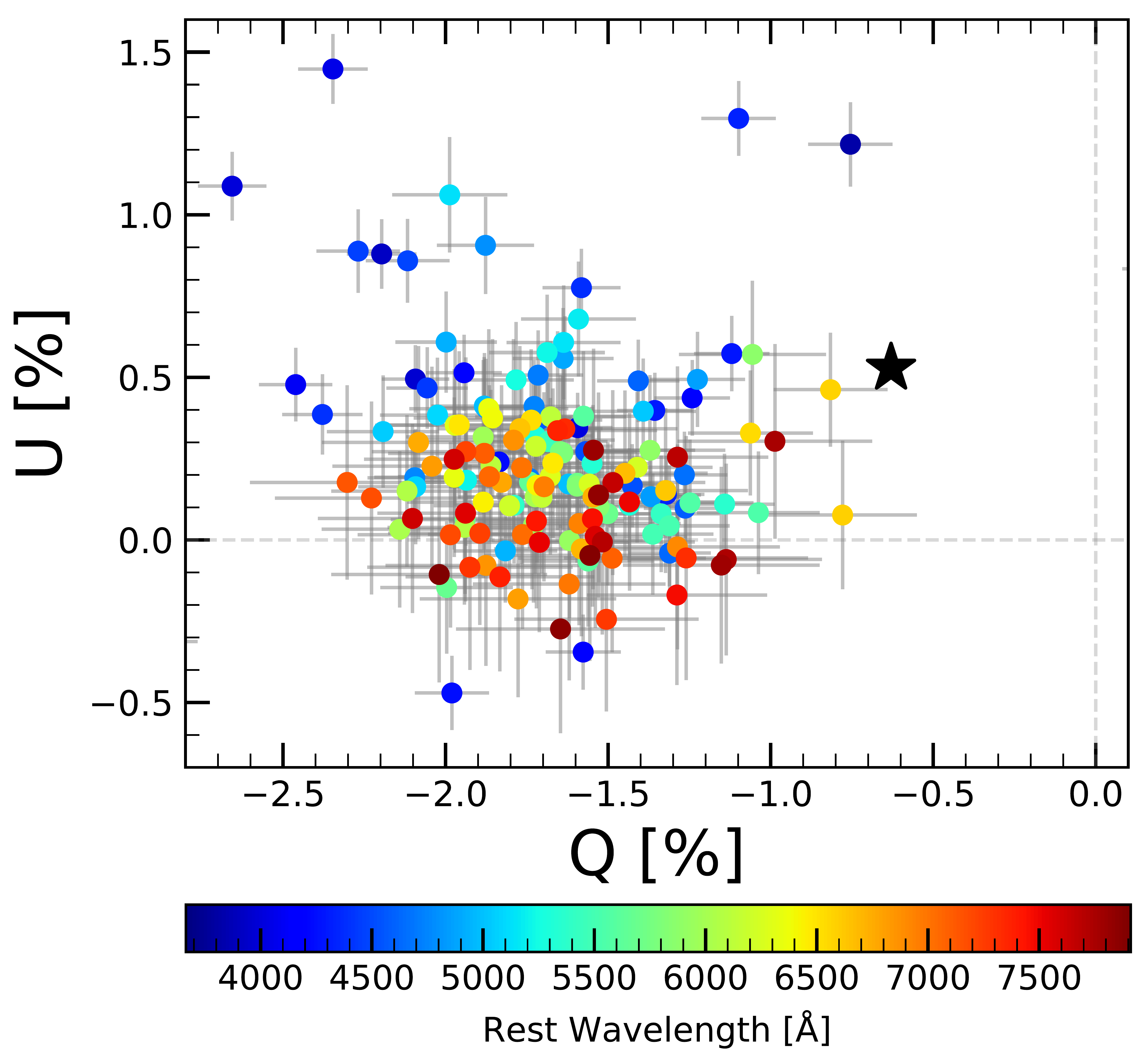}
\caption{Polarization of SN 2020uem in the $Q-U$ diagram, as binned by 20 pixels. The black star symbol is the estimated ISP which is averaged over the wavelength coverage. The color of the points corresponds to the wavelength shown in the bottom colorbar. Note that $P$ and $\theta$ are defined as follows: $P = \sqrt{Q^{2} + U^{2}}$, and $\theta = 1/2 \arctan\left( U/Q \right)$.}
\label{fig:3}
\end{figure}

In Figure \ref{fig:3}, we plot the non-ISP-subtracted Stokes parameters of SN 2020uem on the Q-U plane. In the Q-U plane, the data points are localized in a small region at a certain angle ($\sim 100$ degree) as measured from the ISP. Namely, the polarization angle and degree are overall constant as a function of wavelength. This suggests that SN 2020uem has an asymmetric structure with a preferential direction, i.e., axisymmetric configuration. Note that the scatter of the data points on the bluer wavelength is due to the low S/N.

\subsection{Time Evolution of Polarization} \label{sec:3.3}

\begin{figure}
\epsscale{1.17}
\plotone{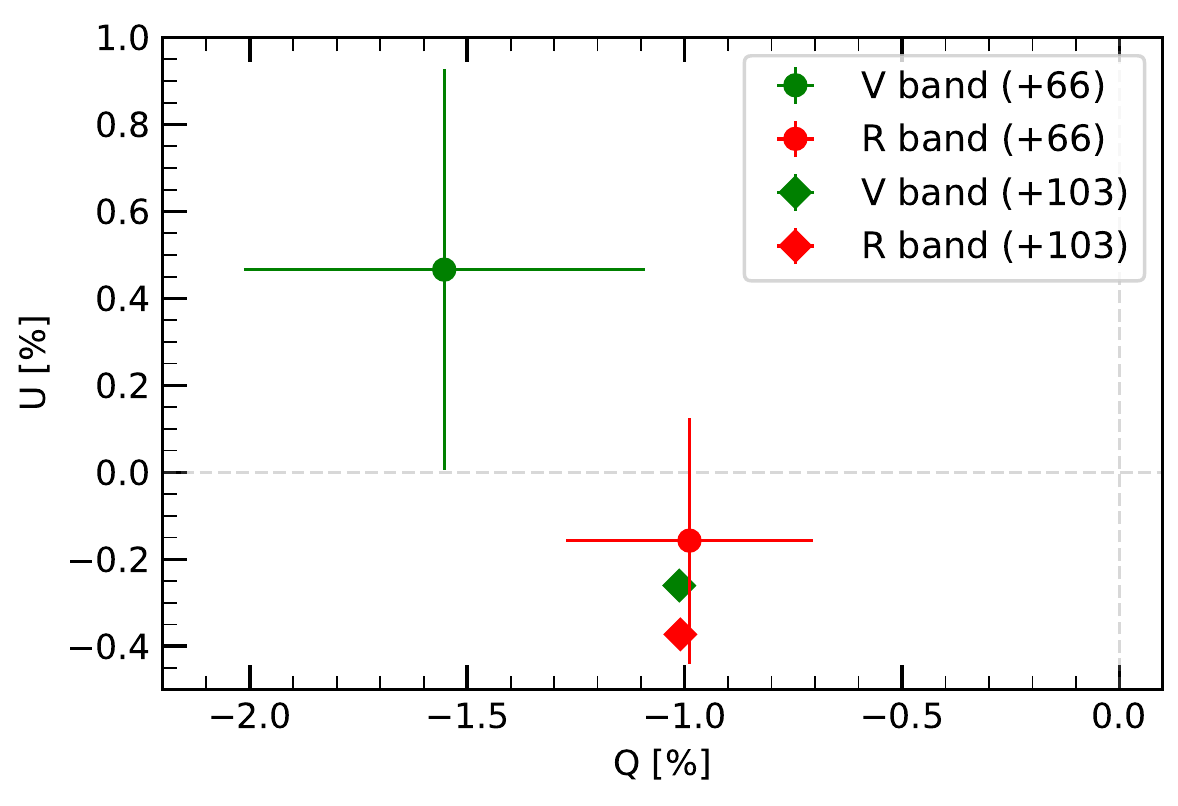}
\caption{Time evolution of the ISP-subtracted broad-band polarization of SN 2020uem in the Q-U diagram. The data at $+66$ days are obtained by the imaging polarization. The data at $+103$ days are reconstructed by the results of the spectropolarimetry. Note that $P$ and $\theta$ are defined as follows: $P = \sqrt{Q^{2} + U^{2}}$, and $\theta = 1/2 \arctan\left( U/Q \right)$.}
\label{fig:4}
\end{figure}

Figure \ref{fig:4} shows the time evolution of the Stokes parameters for SN 2020uem from $+66$ days to $+103$ days on the Q-U plane. The data at $+66$ days is obtained by the imaging polarimetry with Dipol-2 (see Section \ref{sec:2.2}). For the polarization at $+103$ days, we reconstructed the broad-band Stokes parameters by integrating the FOCAS spectropolarimetry data within the V and R bandwidths of Dipol-2. As the ISP for the imaging polarimetry, we use the ISP estimated from the spectropolarimetry.

Both the polarization degree and the position angle do not evolve significantly during the $\sim 40$ days. The polarization degree and angle at two epochs are consistent within the 2-sigma error. This result suggests that SN 2020uem keeps a continuous interaction without evolution in its characteristic shape, i.e., SN 2020uem is powered by an interaction with a globally-asymmetric CSM with a well-defined axis of symmetry, rather than a CSM with a randomly-distributed clumpy structure that will lead to frequent changes in the polarization angle.  
Note that we do not reject the possibility that local fluctuation, e.g., a small-scale clumpy structure, is embedded within the global structure we trace with the polarization.

\section{CSM Geometry} \label{sec:4}

We found that SN 2020uem shows high polarization of $P \sim 1.0-1.5\%$ without wavelength dependence. Besides, the polarization degree and angle keep constant. These results qualitatively indicate a strong interaction with an asymmetric CSM with a well-defined axis of symmetry. Here, we discuss the CSM geometry more quantitatively. Assuming an elliptical CSM with a major axis length of $a$ and a minor axis length of $b$ for simplicity (see Appendix \ref{sec:appendixA}), we estimate the linear polarization by electron scattering using the similar method presented by \citet{Matsumoto2018MNRAS}. 

The observed polarization degree ($P(\theta, b/a)$) as a function of the viewing angle $(\theta)$ and the axis ratio $(b/a)$ is given as follows:
\begin{align}
    P(\theta, b/a) &= P_{\rm max} (\tau_{\rm max}, b/a) \Pi_{0}(\theta) \\
    &= P_{\rm max} (\tau_{\rm max}, b/a) \times 2\left(1 - \frac{b}{a}\right)\sin^{2}\theta,
\end{align}
where $P_{\rm max}$ is the intrinsic polarization degree \citep[see, ][]{Hoflich1991AA} and $\Pi_{0}(\theta)$ is the viewing angle dependence \citep[see, ][]{Brown1977AA}. By using the result of \citet{Hoflich1991AA} (their Fig. 7), we estimate the degrees of polarization for various axis ratios and viewing angles.

\begin{figure}
\epsscale{1.17}
\plotone{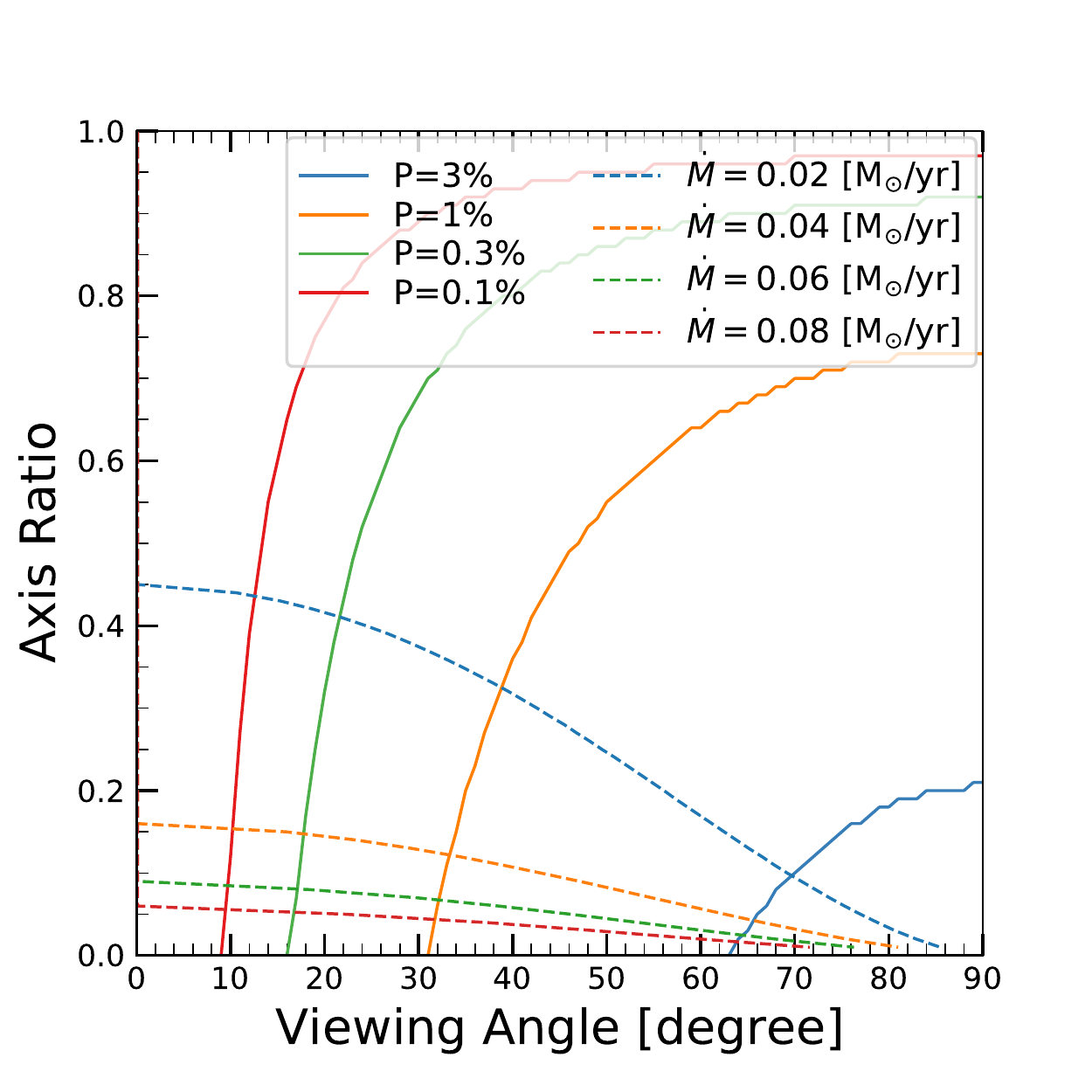}
\caption{Estimated polarization degree as a function of the axis ratio and viewing angle. The blue, orange, green, and red solid lines correspond to the polarization degrees of $3$, $1$, $0.3$, and $0.1\%$, respectively. The dashed lines show the limits above which the optical depth along the line of sight exceeds unity. The blue, orange, green, and red lines correspond to the mass-loss rates of $0.02$, $0.04$, $0.06$, and $0.08 {\rm ~M_{\odot}~yr^{-1}}$, respectively.}
\label{fig:5}
\end{figure}

In Figure \ref{fig:5}, we plot the contours of the axis ratio and the viewing angle which result in the same degree of polarization in the elliptical CSM. To obtain the high polarization degree of $\gtrsim 1\%$, this model requires that the axis ratio is less than $\sim 0.7$ and the viewing angle is larger than $\sim 30^{\circ}$, i.e., the acceptable solution in terms of the polarization degree is the lower part of the orange solid line in Figure \ref{fig:5}.

In Figure \ref{fig:5}, we overplot the constraint based on the optical depth along the line of sight, i.e., the formation of the photosphere, assuming the following `spherical' mass-loss rates; $\dot{M}_{\rm sphere} = 0.02, 0.04, 0.06$, and $0.08 {\rm ~M_{\odot}}$ yr$^{-1}$. It is the same range of the mass-loss rates adopted in section \ref{sec:5}. The dotted lines in Figure \ref{fig:5} show the boundaries where the optical depth is unity for different mass-loss rates. Taking the axis ratio into account, we modify the spherically-symmetric mass-loss rates to elliptical mass-loss rates (see also Appendix \ref{sec:appendixA}).

The early-phase spectra of SN 2020uem exhibit characteristic broad features in its continuum, which likely come from Fe-peak elements, originated either in the ejecta or the shocked CSM (see Paper I). Irrespective the origin of the emission, it requires that the (unshocked) CSM is sufficiently transparent so that the Fe emission is not hidden. Namely, it is rational to assume that the optical depth along the viewing angle is less than 1. Imposing this spectral constraint together with the polarization constraint, the feasible ranges of the axis ratio and viewing angle are limited to $\lesssim 0.3$ and $\sim 30-50^{\circ}$, respectively.

The results require an extremely flat CSM if the elliptical shape is assumed. A straightforward interpretation is a disk- or torus-shaped CSM. Indeed, in the case of a torus CSM, the optical-depth constraint can be automatically satisfied (depending on the viewing direction), since the direct emission coming either from the SN ejecta or the shocked CSM is observable from a relatively pole-on direction. In addition, the high polarization is expected in such aspherical geometry. Therefore, we conclude that SN 2020uem has a disk or torus CSM.

\section{Light-Curve Modeling} \label{sec:5}

In this section, we present a light-curve model for the radiation from an interaction between the SN ejecta and the CSM disk. Here, energy inputs by other processes (e.g., $^{56}$Co decay within the SN ejecta) are not included as the interaction power should overwhelm any other energy sources. We calculate the bolometric luminosity from the interaction shock between the ejecta and the CSM disk ($L_{\rm{disk}}$) using a semi-analytical model. The model is a modified version of a model proposed by \citet{Moriya2013MNRAS_b} and \citet{Nagao2020MNRAS} including diffusion effects within the shock and the CSM on the luminosity evolution.

In Paper I, we have already computed the quasi-bolometric luminosity ($L_{\rm opt}$) using only optical light curves. \citet{Inserra2016MNRAS} shows that the percentage of the bolometric flux in the optical wavelengths for SNe IIn/Ia-CSM is roughly $\sim 80\%$. Using the results, we reconstruct the bolometric luminosity ($L_{\rm bol}$) from the quasi-bolometric luminosity, i.e., $L_{\rm bol} = L_{\rm opt}/0.8$.

The main focus in the analysis here is to investigate whether the white-dwarf progenitor scenario is feasible for SN 2020uem in view of its light curve, and if so, to derive the nature of the CSM based on this scenario. It should be emphasized that we do not attempt to reject a massive star origin. Discriminating these two scenarios is beyond the scope of the present work.

\subsection{SN Properties} \label{sec:5.1}

Assuming an explosion of a WD, we set the mass and kinetic energy of the SN ejecta as  $M_{\rm{ej}}=1.4 {\rm ~M_{\odot}}$ and $E_{\rm{ej}}=1.0 \times 10^{51} {\rm ~erg}$. Considering the results of hydrodynamics simulations \citep[e.g.,][]{Matzner1999ApJ}, the density profile of the SN ejecta is assumed to follow the double power-law distribution \citep[e.g.,][]{Moriya2013MNRAS_b}, which is described as folows:
\begin{eqnarray}
\label{eq:rho_ej}
\widetilde{\rho}_{\rm{ej}} (v_{\rm{ej}},t) = 
\left\{
    \begin{array}{l}
      \frac{1}{4 \pi (n-\delta)} \frac{[2(5-\delta)(n-5) E_{\rm{ej}}]^{\frac{n-3}{2}}}{[(3-\delta)(n-3) M_{\rm{ej}}]^{\frac{n-5}{2}}} t^{-3} v_{\rm{ej}}^{-n} \;\;\;\; (v_{\rm{ej}} > v_{t})\\
      \frac{1}{4 \pi (n-\delta)} \frac{[2(5-\delta)(n-5) E_{\rm{ej}}]^{\frac{\delta-3}{2}}}{[(3-\delta)(n-3) M_{\rm{ej}}]^{\frac{\delta-5}{2}}} t^{-3} v_{\rm{ej}}^{-\delta} \;\;\;\; (v_{\rm{ej}} < v_{t})
    \end{array}
  ,\right.
\end{eqnarray}
where $v_{\rm{ej}}(r,t)= r/t$ is the ejecta velocity and $v_{t}$ is a parameter defined as follows:
\begin{equation}
    v_{t} = \left[ \frac{2(5-\delta)(n-5)E_{\rm{ej}}}{(3-\delta)(n-3)M_{\rm{ej}}} \right]^{\frac{1}{2}}.
\end{equation}
Hereafter, the ejecta density structure is frequently referred to as $\rho_{\rm{ej}} (r,t)$, using the independent variable $r$ instead of $v_{\rm{ej}}$. Note that the power-low index of $n$ does not affect the light curve behavior significantly. Here, we assume that $n=10$ and $\delta = 1$, which are typical values for SNe Ia \citep[e.g.,][]{Matzner1999ApJ}.

\subsection{CSM Properties} \label{sec:5.2}

As an initial distribution of the CSM, we adopt the disk structure with a half-opening angle, $\theta_{\rm disk}$. We adopt the values of $\theta_{\rm disk}=15^{\circ}, 30^{\circ}, 45^{\circ}, 60^{\circ}, 75^{\circ},$ and $90^{\circ}$. We use the radial distribution of the CSM as 
\begin{equation}
    \rho_{\rm{CSM}}(r) = \frac{\dot{M}_{\rm disk}}{\omega_{\rm disk} \times 4\pi v_{\rm{CSM}}r^{2}} = \frac{\widetilde{\dot{M}}_{\rm sphere}}{4\pi v_{\rm{CSM}}r^{2}} = D r^{-2},
\end{equation}
where $v_{\rm{CSM}}$ is the velocity of the CSM, which is estimated as $v_{\rm{CSM}} = 100 {\rm ~km~s^{-1}}$ from the high-dispersion spectroscopic observation (see Paper I), and $D$ is a density scale. Here, $\dot{M}_{\rm disk}$ is the net mass-loss rate for the CSM disk, while $\widetilde{\dot{M}}_{\rm sphere}$ is an isotropic equivalent for the spherically-distributed CSM with the same radial density distribution of the disk CSM. Thus, $\widetilde{\dot{M}}_{\rm sphere}$ and $\dot{M}_{\rm disk}$ are connected as follows:
\begin{equation}
\widetilde{\dot{M}}_{\rm sphere} = \frac{\dot{M}_{\rm disk}}{\omega_{\rm disk}}.
\end{equation}
where,
\begin{equation}
\omega_{\rm{disk}} = \frac{\Omega_{\rm{disk}}}{4 \pi} = \sin{\theta_{\rm disk}}.
\end{equation}
Here, $\Omega_{\rm{disk}}$ is the solid angle of the CSM. The inner radius of the disk is assumed to be $R_{\rm{CSM, in}}$. We adopt the values of $R_{\rm{CSM, in}} = 1.0 \times 10^{15} {\rm ~cm}$, $2.0 \times 10^{15}{\rm ~cm}$ and $3.0 \times 10^{15}{\rm ~cm}$. In the outer region, the CSM is assumed to be distributed infinitely. We compute the bolometric light curves with the disk mass-loss rates from $\dot{M}_{\rm disk}=1 \times 10^{-2} {\rm ~M_{\odot}~yr^{-1}}$ to $1 \times 10^{-1} {\rm ~M_{\odot}~yr^{-1}}$ with an increment of $1 \times 10^{-2} {\rm ~M_{\odot}~yr^{-1}}$.

\subsection{Bolometric Luminosity from The Shock} \label{sec:5.3}

Assuming the physically-thin shocked shell compared to its radius, we calculate the evolution of the shocked shell from the equation of motion as follows:
\begin{eqnarray}
\label{eq:eom}
    M_{\rm{sh}} (t) \frac{{\rm{d}}v_{\rm{sh}}(t)}{ {\rm{d}}t} &=& 4 \pi r_{\rm{sh}}^{2}(t) \Bigl[ \rho_{\rm{ej}} (r_{\rm{sh}}(t),t) \left( v_{\rm{ej}}(r_{\rm{sh}}(t),t)-v_{\rm{sh}}(t) \right)^{2} \nonumber\\
    && - \rho_{\rm{CSM}} (r_{\rm{sh}}(t)) \left( v_{\rm{sh}}(t)-v_{\rm{CSM}} \right)^{2} \Bigr],
\end{eqnarray}
where $M_{\rm{sh}}(t)$ is the total mass of the shocked SN ejecta and the shocked CSM at given time, $t$, and $v_{\rm{sh}}(t)$ is the velocity of the shocked shell. Here $M_{\rm{sh}}(t)$ is expressed as follows:
\begin{equation}
    M_{\rm{sh}} (t) = \int_{R_{\rm{CSM, in}}}^{r_{\rm{sh}}(t)} 4 \pi r^{2} \rho_{\rm{CSM}} (r) dr + \int_{r_{\rm{sh}}(t)}^{r_{\rm{ej,max}}(t)} 4 \pi r^{2} \rho_{\rm{ej}} (r,t) dr,
\end{equation}
where $r_{\rm{ej,max}}(t) = v_{\rm{ej,max}} t$, and $v_{\rm{ej,max}}$ is the original velocity of the outermost layer of the SN ejecta before the interaction. Here, we assume $r_{\rm{ej,max}}(t) \gg r_{\rm{sh}}(t)$ at all times. We obtain the values of $r_{\rm{sh}}(t)$ and $v_{\rm{sh}}(t)$ by numerically solving Eq.~\eqref{eq:eom}.

First, we consider light curves for interacting SNe with a spherically symmetric CSM. In the optical thin limit, a fraction of the generated energy, which can escape from the shocked shell as radiation \citep[see, e.g.,][]{Moriya2013MNRAS_b, Nagao2020MNRAS}, is described as follows:
\begin{equation}
    L_{\rm{sh}}(t) = \varepsilon \frac{{\rm{d}}E_{\rm{kin}}(t)}{{\rm{d}}t} = 2 \pi \epsilon \rho_{\rm{CSM}} (r_{\rm{sh}}(t)) r_{\rm{sh}}^{2}(t) v_{\rm{sh}}^{3}(t),
\end{equation}
where $\varepsilon$ is the conversion efficiency from kinetic energy to radiation, and
\begin{equation}
    {\rm{d}}E_{\rm{kin}}(t) = 4 \pi r_{\rm{sh}}^{2}(t) \left( \frac{1}{2} \rho_{\rm{CSM}} (r_{\rm{sh}}(t)) v_{\rm{sh}}^{2}(t) \right) \rm{d}r.
\end{equation}
In our calculation, we use the efficiency as a free parameter. We adopt a best-fit $\varepsilon$ that minimizes the residuals between the model and the observed data. The residuals are defined as follows: ${\rm residual} = \sum [({\rm model} - {\rm data})/{\rm data}]^{2}$.

When the optical depth in the shocked shell exceeds unity, the optical-depth effects become important. Since the generated photons stay in the shocked shell for diffusion time, their emergence becomes delayed. During this confinement of the photons, their energy is used for accelerating the shocked shell. However, in the end, this additional kinetic energy of the shocked shell comes back to the thermal energy due to the shock interaction in the forward/reverse shock fronts. Thus, the energy of photons is not lost in this process. In this work, we assume that all the generated photons due to the continuing shock interaction escape from the shock in the diffusion time. After the escape from the shock, the photons stay also in the CSM for a while. Thus, their arrival to the observer becomes more delayed, and their energy is not changed also in this diffusion process within the CSM. This is because the photons have already been decoupled from the shocked gas, i.e., they are already after the shock breakout, where the diffusion time is smaller than the dynamical time of the shock. 

We calculate a light curve from the interaction as a superposition of gas shells with photons that are generated at each time \citep[see][]{Nagao2020MNRAS}. The photons generated in a time interval between $t-{\rm{d}}t/2$ and $t+{\rm{d}}t/2$ experience diffusion processes with $\tau_{\rm{diff}}(t)$ as follows:
\begin{eqnarray}
    \tau_{\rm{diff}}(t) &=& \kappa_{\rm{es}} \frac{M_{\rm{sh}}(t)}{4 \pi r_{\rm{sh}}^{2} (t) \Delta R_{\rm{sh}}(t)} \Delta R_{\rm{sh}}(t) + \int_{r_{\rm{sh}}(t)}^{\infty} \kappa_{\rm{es}} \rho_{\rm{CSM}} dr\\ 
    &=& \frac{\kappa_{\rm{es}} M_{\rm{sh}}(t)}{4 \pi r_{\rm{sh}}^{2} (t)} + \frac{\kappa_{\rm{es}}D}{(s-1)}r_{\rm{sh}}^{-(s-1)}(t),
\end{eqnarray}
where $\Delta R_{\rm{sh}}(t)$ is the width of the shocked shell at given time ($t$), and $\kappa_{\rm{es}}$ is the mass scattering coefficient for the ionized gas. Throughout the paper, we adopt the value of $\kappa_{\rm{es}}=0.34 \; \rm{cm}^{2} ~\rm{g}^{-1}$ as opacity in the fully ionized gas mainly composed of hydrogen and helium at the solar metallicity. 

Thus, the light curve from the interaction with the spherical CSM is computed as follows:
\begin{equation}
    L_{\rm{sphere}} (t) = \int_{0}^{t} \frac{L_{\rm{sh}}(t^{\prime}) \rm{d}t^{\prime}}{t_{\rm{diff}}(t^{\prime})} \exp \left( - \frac{t-t^{\prime}}{t_{\rm{diff}}(t^{\prime})} \right),
\end{equation}
where
\begin{equation}
    t_{\rm{diff}} (t) = \frac{\tau_{\rm{diff}}(t) r_{\rm{sh}}(t)}{c}.
\end{equation}

In the case of the interaction with the disk CSM, the CSM is confined into the small region, where the mass-loss rate for the disk ($\dot{M}_{\rm{disk}}$) is defined under the assumption that there is no mass loss except for that responsible to the CSM disk. Thus, the density scale of the CSM disk corresponds to that of the spherical CSM with $\widetilde{\dot{M}}_{\rm{sphere}}=\dot{M}_{\rm{disk}}/\omega_{\rm{disk}}$. Therefore, the bolometric light curves from interacting SNe with the disk CSM are calculated as follows:
\begin{equation}
\label{eq:Ldisk}
    L_{\rm{disk}} (t,\dot{M}_{\rm{disk}}) = \omega_{\rm{disk}} L_{\rm{sphere}} \left(t, \widetilde{\dot{M}}_{\rm{sphere}} \right),
\end{equation}

In this model, we cannot take the multi-dimensional effects, e.g., light-travel time or viewing angle effects, into account. However, compared to the timescale of variation of the light curve ($\gtrsim 10$ days), the light-travel time of this system is sufficiently small ($\lesssim$ a few days) since the typical shocked radius is $\sim 10^{16}{\rm ~cm}$. Therefore, here we neglect the geometrical effects.

\subsection{Results} \label{sec:5.4}

\begin{figure*}
\gridline{\fig{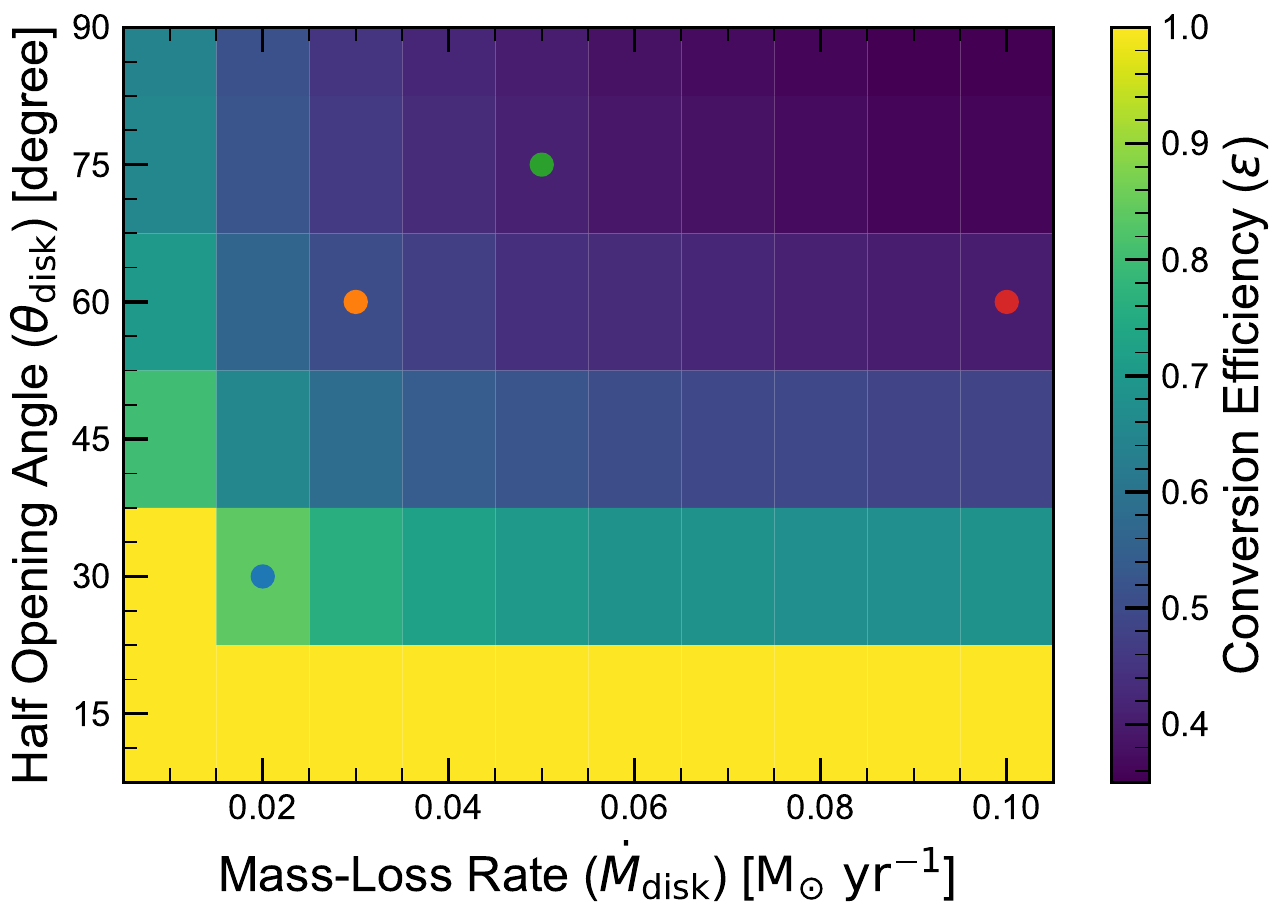}{0.45\textwidth}{(A)}
          \fig{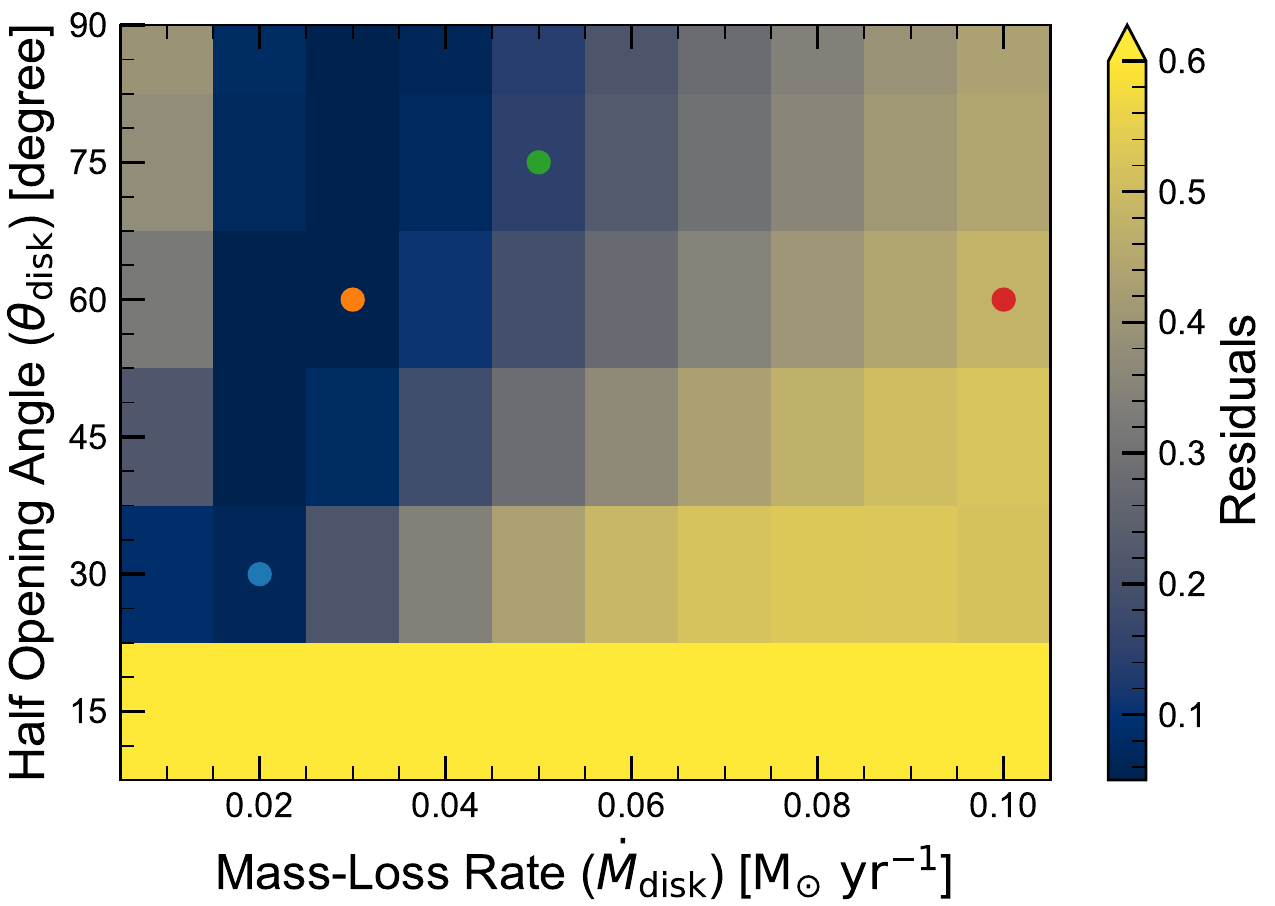}{0.45\textwidth}{(B)}
          }
\gridline{\fig{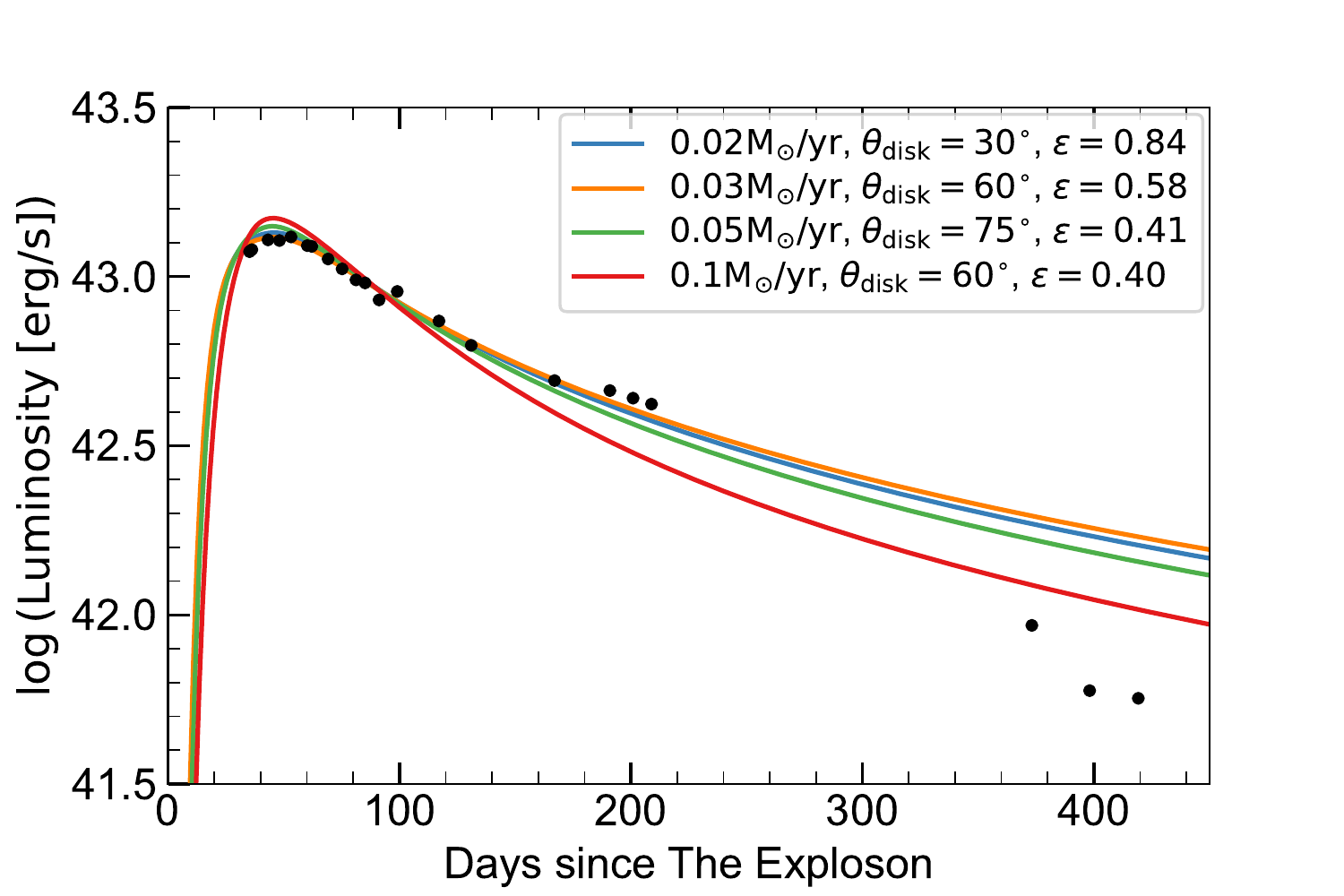}{0.45\textwidth}{(C)}
          \fig{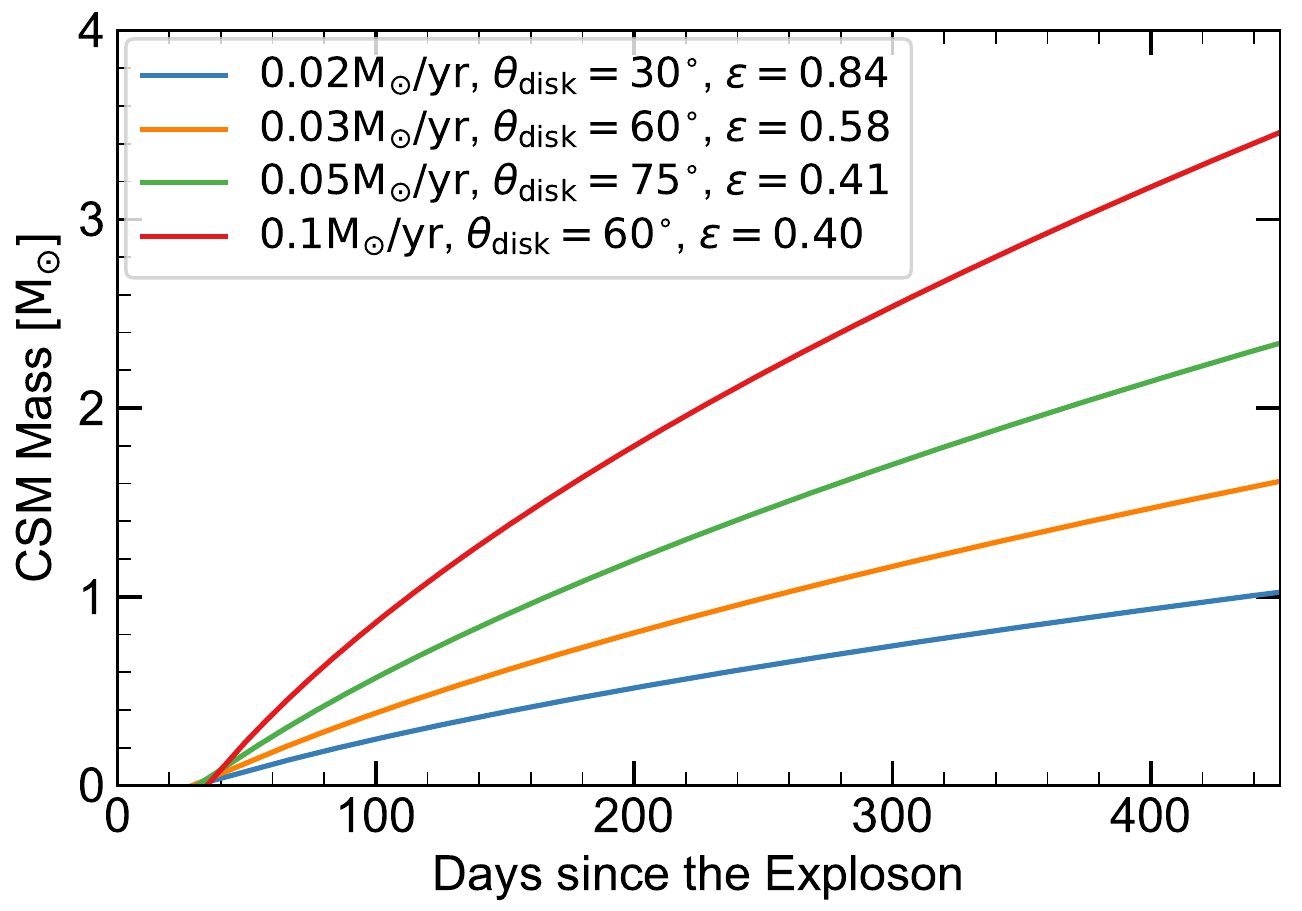}{0.45\textwidth}{(D)}
          }
\caption{The results of the mass-loss rate modeling. Panel (A) shows the best-fit values for the energy conversion efficiency $\varepsilon$ as a function of the mass-loss rate and disk half-opening angle $\theta_{\rm disk}$. The blue, orange, green, and red dots are the models plotted in Panels (C) and (D). In particular, the blue and orange correspond to the best fits models, i.e., the minimum (0.37) and second minimum (0.38) RMS (see also Panel (B)). Panel (B) shows the residuals between the models with the best-fit $\varepsilon$ and observed data, for a given combination of the mass-loss rate and the opening angle. Panel (C) shows the model light curves for some selected models. Panel (D) shows the time evolution of the CSM mass swept by the shock.}
\label{fig:6}
\end{figure*}

In Figure \ref{fig:6}, we show the results of the models with $R_{\rm CSM, in} = 2.0 \times 10^{15}{\rm ~cm}$ (for other parameters, see also Appendix \ref{sec:appendixB}). Note that we do not use the late-phase data ($\gtrsim 350$ days) for the modeling since the light curve shows an acceleration decay (see also Paper I). Panel (A) shows the best-fit values for $\varepsilon$ as a function of the mass-loss rates and opening angles, and panel (B) shows the residuals. Based on the residuals, the feasible parameters of the mass-loss rates and the opening angles are in the range of $0.01 - 0.05 {\rm ~M_{\odot}~yr^{-1}}$ and $30 - 90^{\circ}$, respectively. In a dense CSM, the energy conversion from the kinetic to thermal becomes effective, and then the conversion efficiency is expected to be close to unity \citep[e.g.,][]{Maeda2022ApJ}. In fact, the model shows that the best-fit value for the efficiency is roughly $0.3 - 0.8$. Note that we do not consider the radioactive decay of $^{56}{\rm Ni}$, but the effect is negligible since SN 2020uem shows high luminosity which cannot be powered by the decay (see also Paper I).

In Panel (C), we show the model bolometric light curves for some selected models. We confirm that the models which provide an acceptable fit (judged by visual inspection) are those with small residuals, i.e., the lines colored in blue. The density structure with the CSM power-law index of $-2$ explains the decay rate in the early phase reasonably well, while the late-phase light curve cannot be explained by the same density structure as in the early phase. The result implies that the density structure is changed at $\sim 300$ days between the early and the late phases, or the optical emission becomes suppressed by newly-formed dust. Besides, as yet another scenario, the accelerated decay may be caused by the reverse shock reaching to the bottom of the ejecta, which might happen when the shocked CSM mass becomes comparable to the ejecta mass \citep[e.g.,][]{Svirski2012ApJ, Ofek2014ApJ, Inserra2016MNRAS}. For the first scenario, our estimate on the CSM mass (as derived with the light curve evolution in the one-year time scale) corresponds to the total CSM mass. For the other two scenarios, strictly speaking, our estimate provides only a lower limit while it can still be a measure of the mass budget required for SN 2020uem (see also Paper I). In the future, we hope to discriminate these scenarios by further investigating observational data in the later phase than presented in the present work.

Panel (D) shows the time evolution of the CSM mass swept by the shock. Although we cannot conclude that the late-phase behavior of the light curve is due to the change in CSM density, the acceptable range of the CSM mass is $\sim 0.5 - 4 {\rm ~M_{\odot}}$. The mass range is consistent with previous studies on SNe IIn/Ia-CSM using other models \citep[e.g.,][]{Inserra2016MNRAS}. Unlike SNe IIn which sometimes require a huge mass budget supplied by a massive RSG \citep[e.g.,][]{Smith2011ApJ}, the CSM mass derived for SN 2020uem can be accommodated by a less massive star like an AGB star or a red giant (RG), while an even more massive star, like a super-AGB or RSG can of course provide sufficient mass budget.

The shock velocity is roughly $\sim (3-6) \times 10^{3}{\rm ~km~s^{-1}}$ and the shocked radius is roughly $\sim 10^{16}{\rm cm}$. Considering the radial extent of the CSM and ejecta velocity, the CSM driving the early light curves is expected to have been formed by final activities a few hundred years before the explosion. The result here provides a potentially strong and new constraint on the progenitor scenario of SN 2020uem (see Section \ref{sec:6}).

\section{Discussion and Conclusions} \label{sec:6}

We have performed intensive follow-up observations of an SN IIn/Ia-CSM, SN 2020uem, including photometry, spectroscopy, and polarimetry. In this paper, we focused on the results of the polarimetry and light-curve modeling. Note that the data of photometry and spectroscopy have been presented in Paper I.

We have performed imaging polarimetry at $+66$ days and spectropolarimetry at $+103$ days after the discovery. SN 2020uem keeps showing the high polarization degree of $\sim 1.0-1.5\%$ without wavelength dependence, as well as the constant position angle. The high polarization degree suggests that the CSM around SN 2020uem is confined and the geometry is highly aspherical. Besides, the wavelength- and time-independent polarization implies that SN 2020uem is powered by a strong interaction with a globally-aspherical CSM with a well-defined axis of symmetry. Using a simple model for the polarization, we conclude that SN 2020uem has an equatorial-disk/torus CSM.

Assuming an interaction with an equatorial-disk/torus CSM, we performed light curve modeling with a semi-analytical model and estimated the CSM mass. We found that the mass-loss rate of SN 2020uem is in the range of $0.01 - 0.05 {\rm ~M_{\odot}~yr^{-1}}$. The estimated CSM mass is $0.5-4 {\rm ~M_{\odot}}$, which can be accommodated by an AGB star or an RG star. Besides, the accelerated decay of the light curve indicates that the CSM might have been formed in $\sim {\rm a~few}\times 100$ years before the explosion. Therefore, the DD scenario can be rejected, and a binary system including an AGB or an RG star is possible for the progenitor system for which a binary system with a WD can be naturally considered. Our analysis does not reject a massive super-AGB or an RSG as a mass budget for the formation of the CSM, which is still possible to form a binary with a WD and indeed considered as an extreme case of SNe Ia-CSM \citep{Jerkstrand2020Science}. In the SD scenario, the equatorial-disk/torus CSM may be formed via Roche-lobe overflow. However, the timescale of stellar evolution makes it difficult to form such a massive and packed CSM within a scale of $10^{16}{\rm ~cm}$.

Hence, as an explosion mechanism, we suggest a stellar merger scenario via a CE evolution, i.e., the CD scenario \citep{Livio2003ApJ, Kashi2011MNRAS} and its variant \citep{Jerkstrand2020Science}. In these scenarios, the mass ejection in the CE phase is more likely confined in the radial direction than in the polar direction \citep[e.g.,][]{Kashi2011MNRAS,Iacon2019MNRAS}. This is consistent with the equatorial-disk/torus CSM. Besides, \citet{Soker2013MNRAS} suggested that a CE between a WD and an AGB star can form massive CSM within $\sim 10^{16}{\rm ~cm}$. Therefore, as the possible origin of SN 2020uem, we propose a thermonuclear explosion within a dense CSM triggered by a stellar merger including an RG or AGB star (or perhaps even an RSG or a super AGB star).

The maximum V-band magnitude of SN 2020uem in our observation period is $\sim -19.5{\rm ~mag}$ (see Paper I), and the observed light curves are well explained by the interaction model (see Section \ref{sec:5}). However, we missed the rising phase. Indeed, the initial g- and r-band data obtained by the ZTF survey \footnote{see, e.g., \url{https://alerce.online/object/ZTF20accmutv}}, taken about $\sim 20$ days earlier than our light curves, show brighter magnitudes than the peak of the interaction light curves \citep[see Paper I and light curves provided by ZTF Brokers, e.g., ALeRCE;][]{Forster2021AJ}. This result indicates that the peak magnitude of the underlying SN likely reached to $\sim -20 {\rm ~mag}$. This luminosity fits a bright SN Ia, e.g., 91T-like SNe  \citep{Taubenberger2017SNhandbook}, which also show spectral similarities to SN 2020uem and SNe IIn/Ia-CSM (see also Paper I).

While the well-observed sample of SNe IIn/Ia-CSM is increasing, including this work that provides one of the most massive data sets for SNe IIn/Ia-CSM, it is still very limited. In order to have more advanced and robust discussion, it is necessary to improve the sample of SNe IIn/Ia-CSM. Besides, it is also important to perform late-phase observation, especially spectroscopy. After the CSM interaction is over, the photosphere will recede and the innermost region can become transparent. The late-phase spectra in the post-interaction phase may reflect the information on the progenitor and explosion mechanism of SNe IIn/Ia-CSM. Such data are expected to play a key role in revealing the nature of SNe IIn/Ia-CSM, as demonstrated by \citet{Jerkstrand2020Science}. For a deeper understanding of SNe IIn/Ia-CSM, it is essential to obtain a comprehensive picture from the underlying SN explosion to the overlying CSM environment through continuous observations from the early to late phase of SNe IIn/Ia-CSM.

\begin{acknowledgments}
This research is based on observations obtained at the Subaru Telescope (S20B-056) operated by the Astronomical Observatory of Japan, and at the Tohoku T60 telescope operated by the Tohoku University Haleakala Observatories. The authors thank the staff of the Subaru Telescope and the Tohoku T60 telescope for excellent support in the observations. In particular, the authors thank Andrei Berdyugin for providing the imaging polarimetry data. Besides, we thank the KASTOR (Kanata And Seimei Transient Observation Regime) team, the Subaru/SWIMS (Simultaneous-color Wide-field Infrared Multi-object Spectrograph) developer team, and the TriCCS developer team for obtaining the data used to construct the bolometric luminosity. The authors also thank Daichi Hiramatsu, Takashi J. Moriya, and Nozomu Tominaga for valuable discussion. The authors also thank `1st Finland-Japan bilateral meeting on extragalactic transients', which gave us a good opportunity to discuss this object. K.U. acknowledges financial support from Grant-in-Aid for the Japan Society for the Promotion of Science (JSPS) Fellows (22J22705). K.U. also acknowledges financial support from AY2022 DoGS Overseas Travel Support, Kyoto University. T.N. is funded by the Academy of Finland project 328898. T.N. acknowledges the financial support by the mobility program of the Finnish Centre for Astronomy with ESO (FINCA). K.M. acknowledges support from JSPS KAKENHI grant Nos. JP18H05223, JP20H00174, and JP20H04737. The work is partly supported by the JSPS Open Partnership Bilateral Joint Research Project between Japan and Finland (JPJSBP120229923). H.K. was funded by the Academy of Finland projects 324504 and 328898.

\facilities{Subaru}
\software{IRAF, Astropy \citep{astropy13,astropy18}}

\end{acknowledgments}

\appendix

\section{Optical Depth in the Elliptical CSM} \label{sec:appendixA}

\begin{figure}
\epsscale{1.17}
\plotone{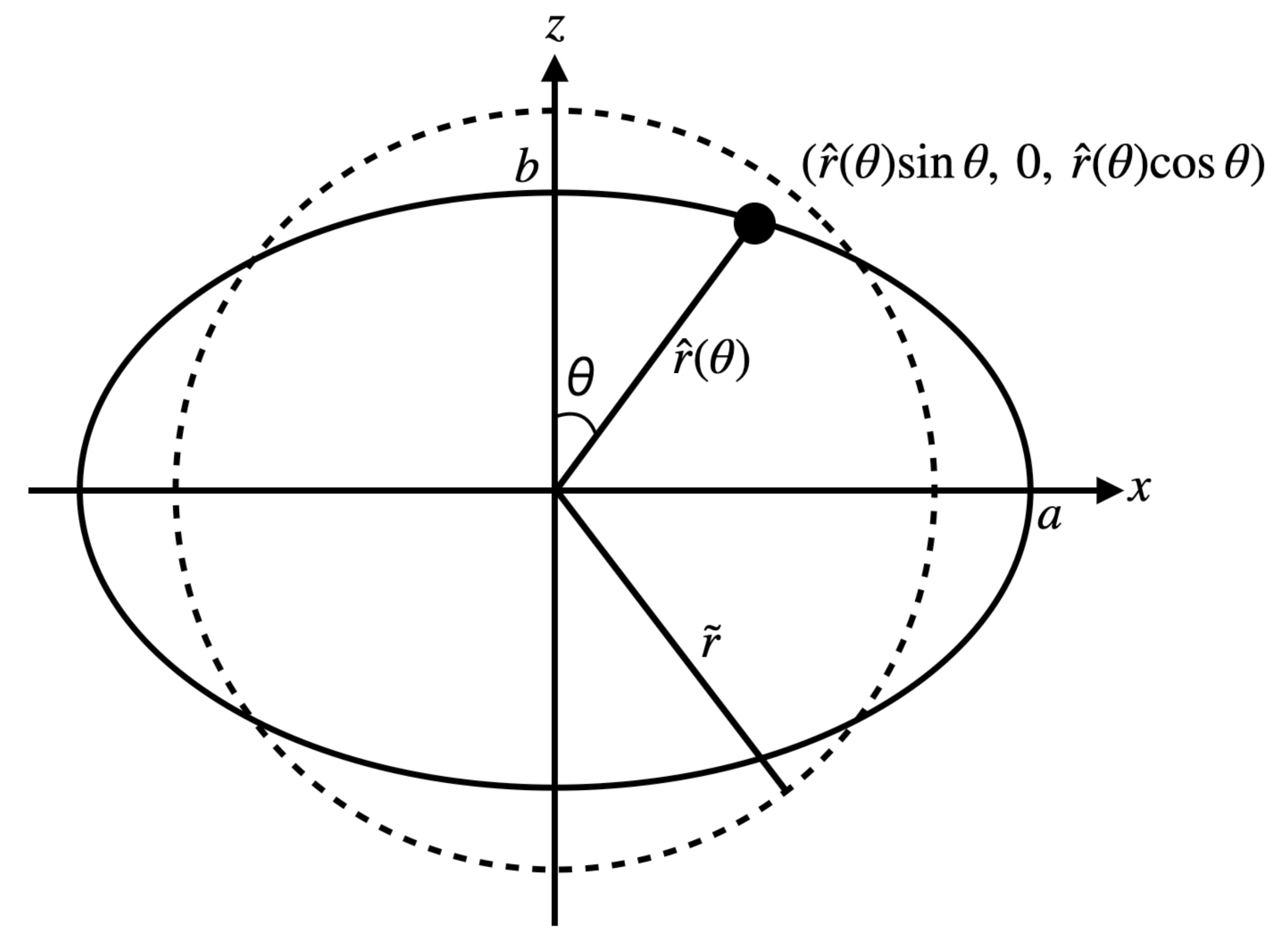}
\caption{The configuration of the calculation of the optical depth in the elliptical CSM.}
\label{fig:app1}
\end{figure}

Assuming constant spherical mass-loss rates ($\dot{M}_{\rm sphere}$), the optical depth in a spherical CSM ($\tau_{\rm sphere}$) is described as follows:
\begin{align}
    \tau_{\rm sphere} &= \int_{r}^{\infty} \kappa_{\rm es}\rho dr = \frac{\kappa_{\rm es}\dot{M}_{\rm sphere}}{4\pi v_{\rm CSM}} \frac{1}{r}.
\end{align}
For an elliptical CSM (see Figure \ref{fig:app1}), the position in the ellipsoid ($\hat{r}$) is given as follows:
\begin{align}
    \hat{r}(\theta) = \left(\frac{a^2 b^2}{b^{2} \sin^{2} \theta + a^{2} \cos^{2} \theta}\right)^{\frac{1}{2}},
\end{align}
where $\theta$ is an angle with respect to the z-axis, and $a$ and $b$ are the major and minor axis length, respectively. The radius of a sphere with the same volume as the ellipsoid is denoted by $\tilde{r}$. The radius is given as follows:
\begin{equation}
    \tilde{r} = (a^2 b)^{\frac{1}{3}}.
\end{equation}
Here, we assume that $\dot{M}_{\rm sphere}$ and the elliptical mass-loss rates ($\dot{M}_{\rm elliptical}$) are connected as follows:
\begin{equation}
    \dot{M}_{\rm elliptical}(\theta) = \dot{M}_{\rm sphere}\frac{r(\theta)}{\tilde{r}}.
\end{equation}
Then, we can obtain the optical depth as a function of $\theta$ in the elliptical CSM as follows:
\begin{align}
    \tau_{\rm elliptical} (\theta) &= \frac{\kappa_{\rm es}\dot{M}_{\rm elliptical}}{4\pi v_{\rm CSM}} \frac{1}{r} \\
    &=\frac{\kappa_{\rm es}\dot{M}_{\rm sphere}}{4\pi v_{\rm CSM}} \left(\frac{b}{a}\right)^{\frac{2}{3}} \left[\left( \frac{b}{a} \right)^{2}\sin^{2}\theta + \cos^{2}\theta \right].
\end{align}

\section{Light Curve Modeling for Other Parameter Sets} \label{sec:appendixB}

\begin{figure*}
\gridline{\fig{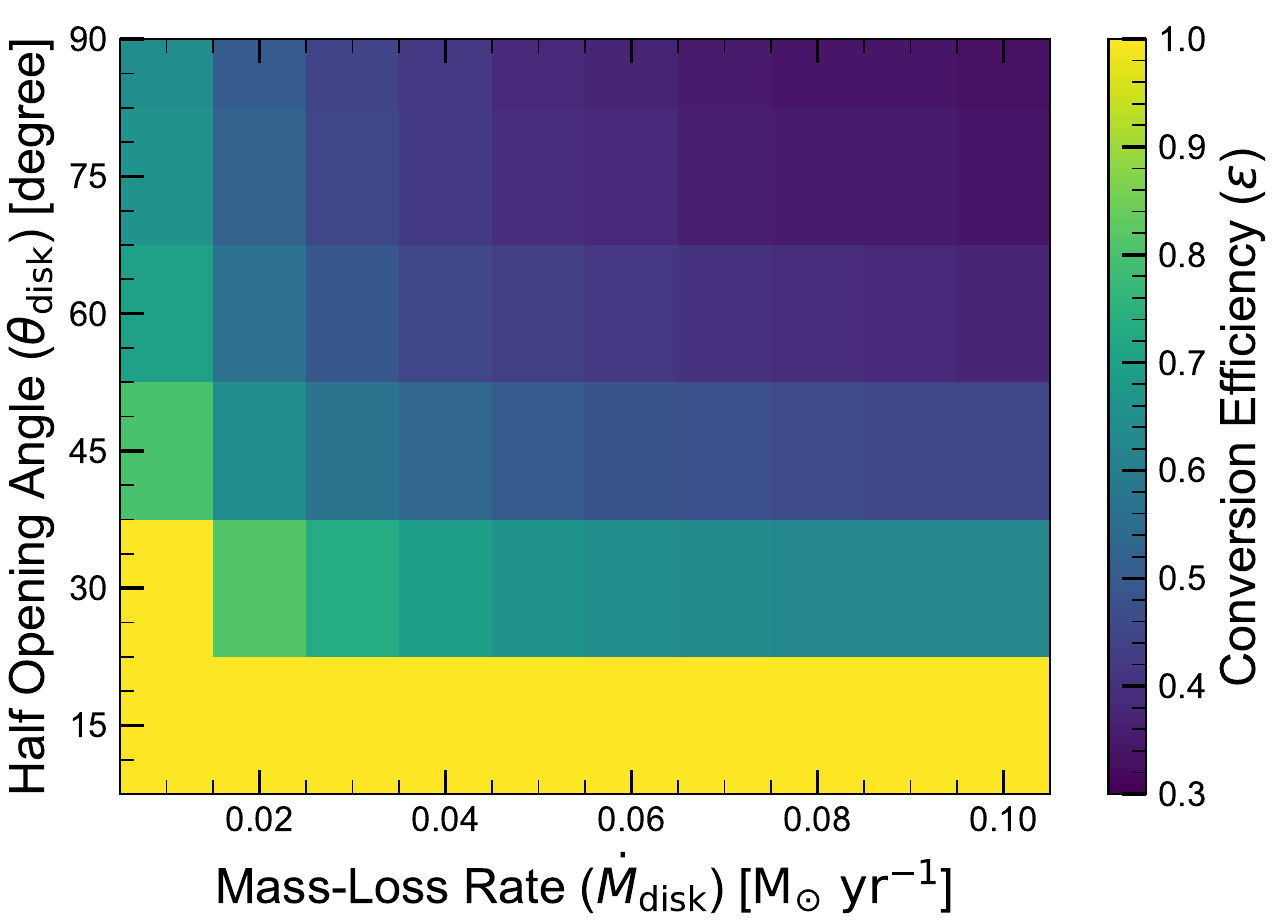}{0.45\textwidth}{(a)}
          \fig{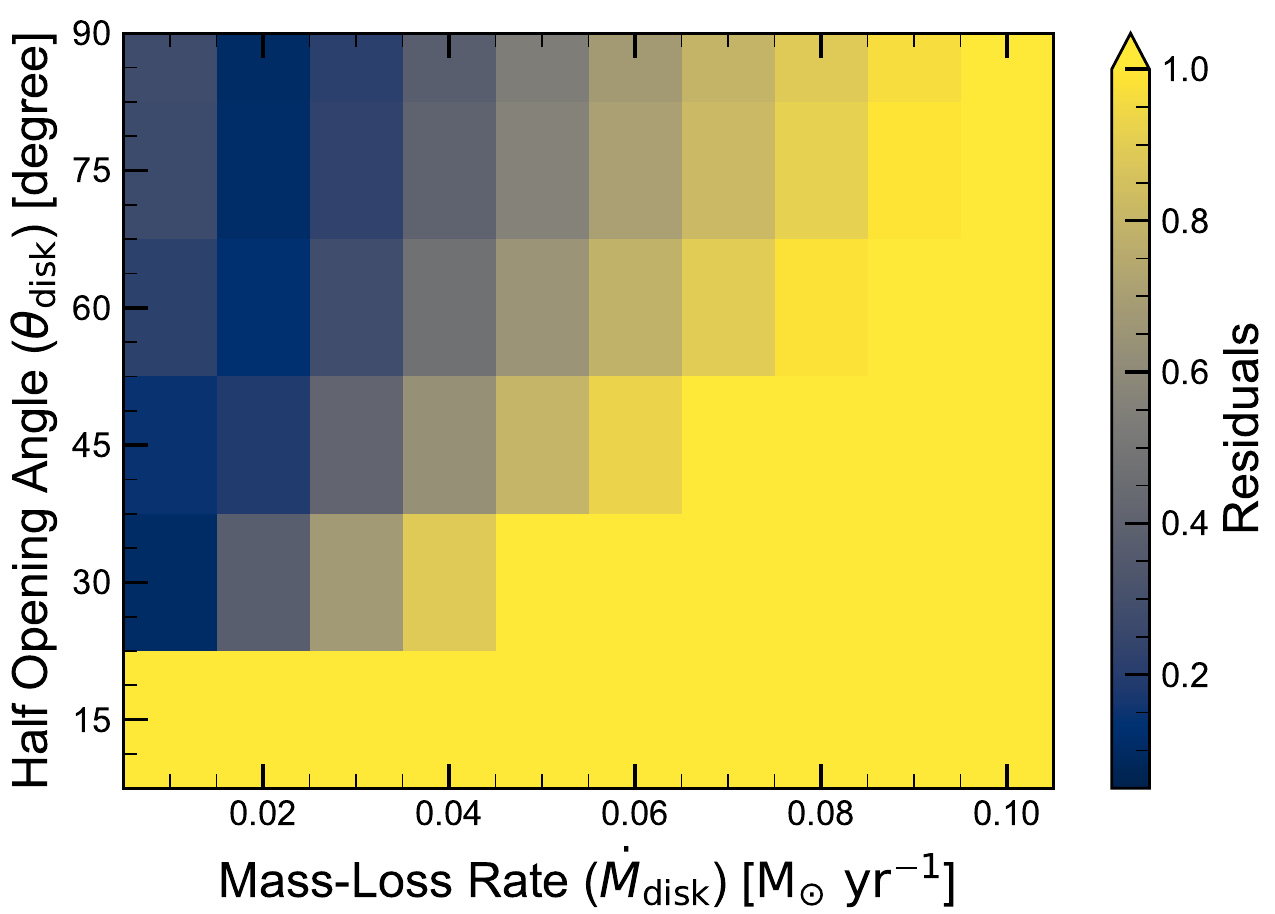}{0.45\textwidth}{(b)}
          }
\gridline{\fig{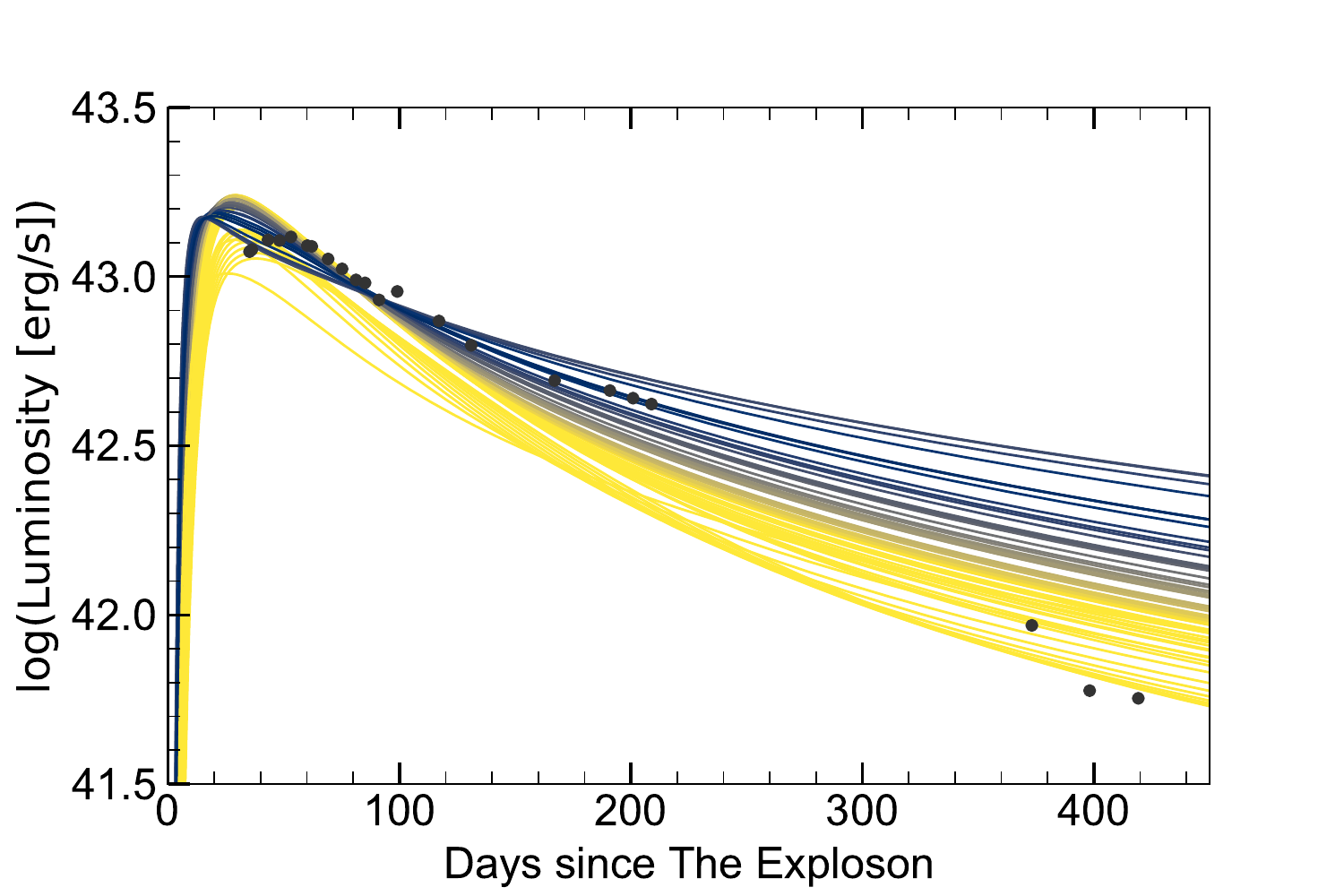}{0.45\textwidth}{(c)}
          \fig{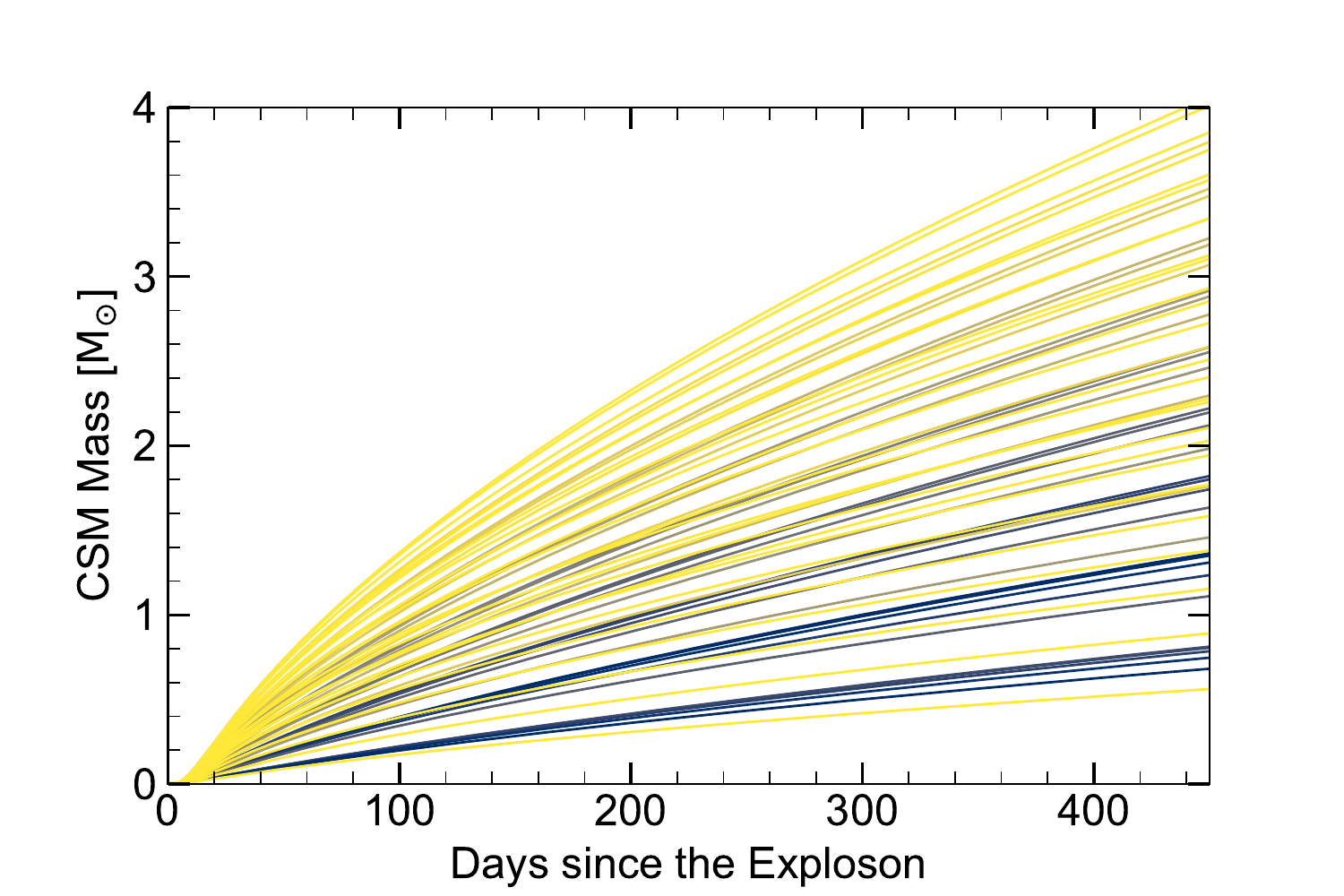}{0.45\textwidth}{(d)}
          }
\caption{The same as Figure \ref{fig:6}, but for $R_{\rm CSM, in} = 1.0\times 10^{15}{\rm cm}$. We plot all the light curve models and CSM mass evolution, as color-coded by the color map found in the residual plot.}
\label{fig:app2}
\end{figure*}

\begin{figure*}
\gridline{\fig{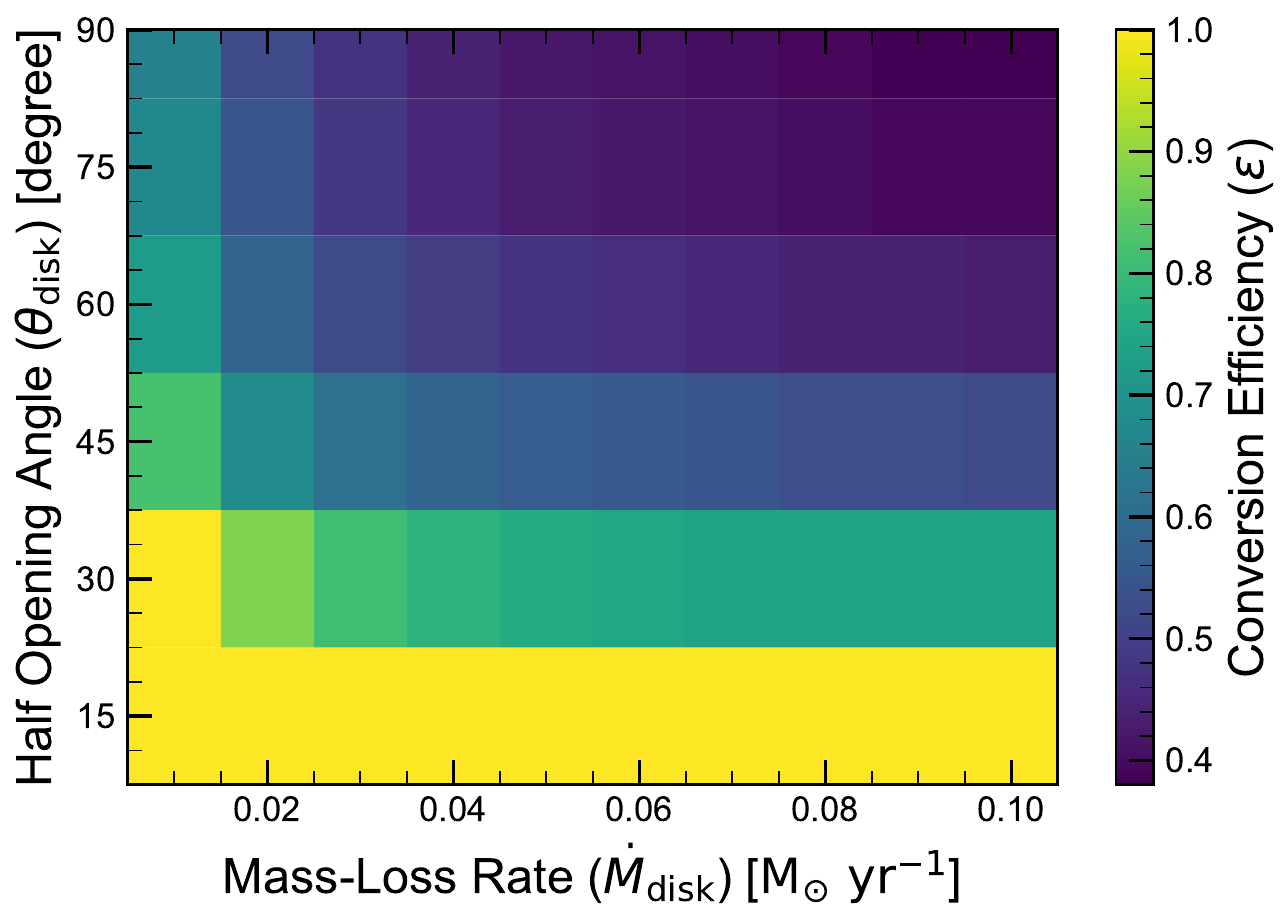}{0.45\textwidth}{(a)}
          \fig{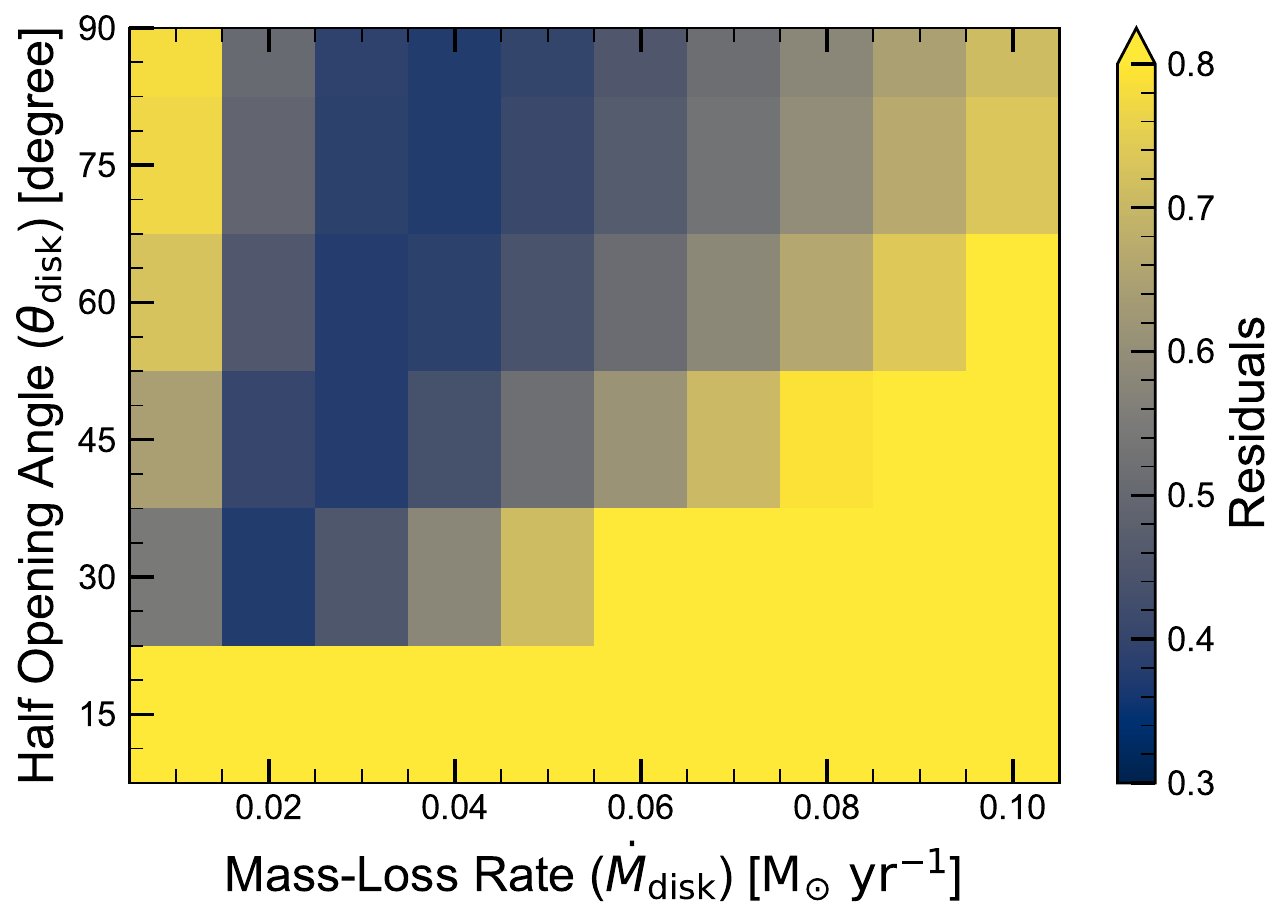}{0.45\textwidth}{(b)}
          }
\gridline{\fig{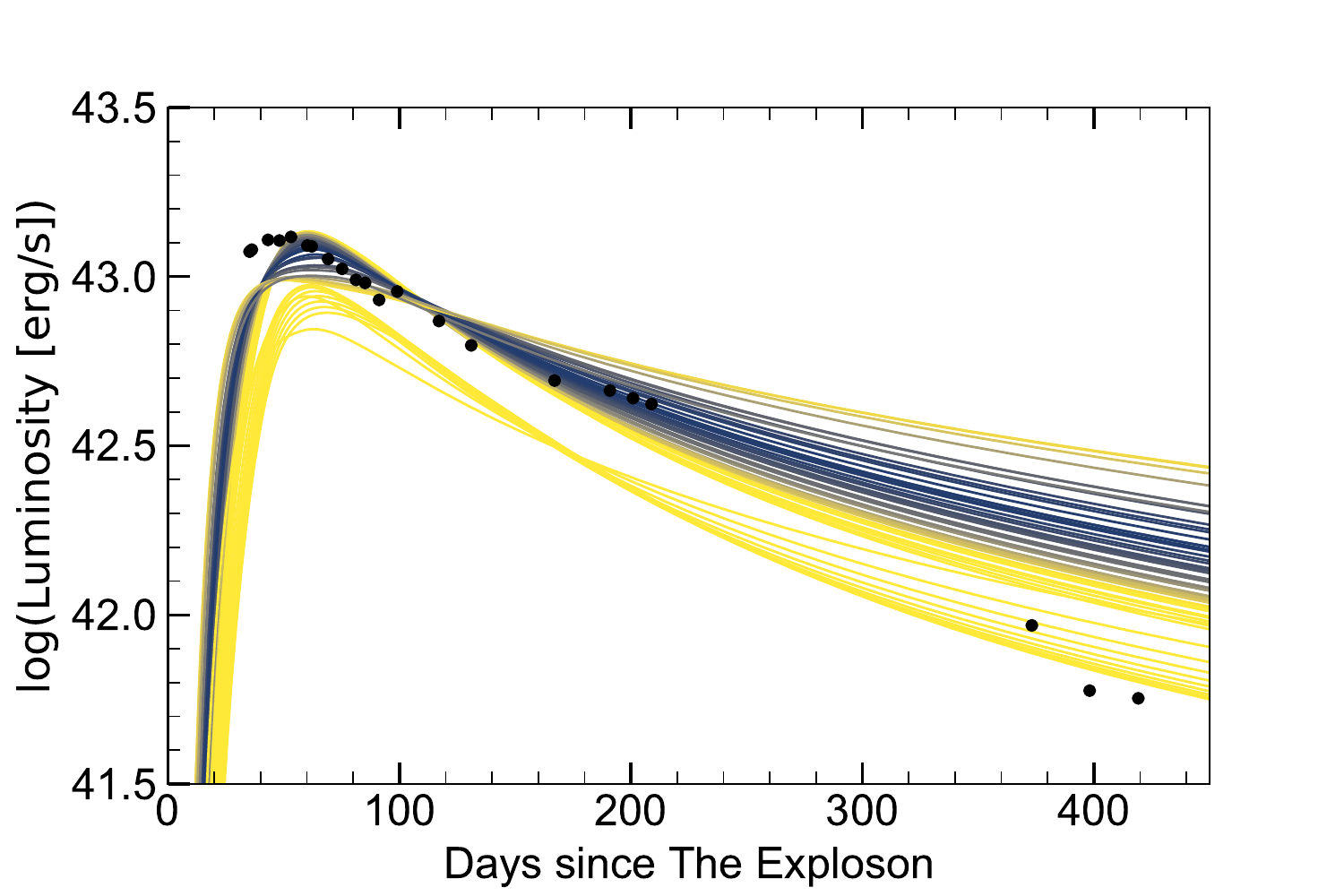}{0.45\textwidth}{(c)}
          \fig{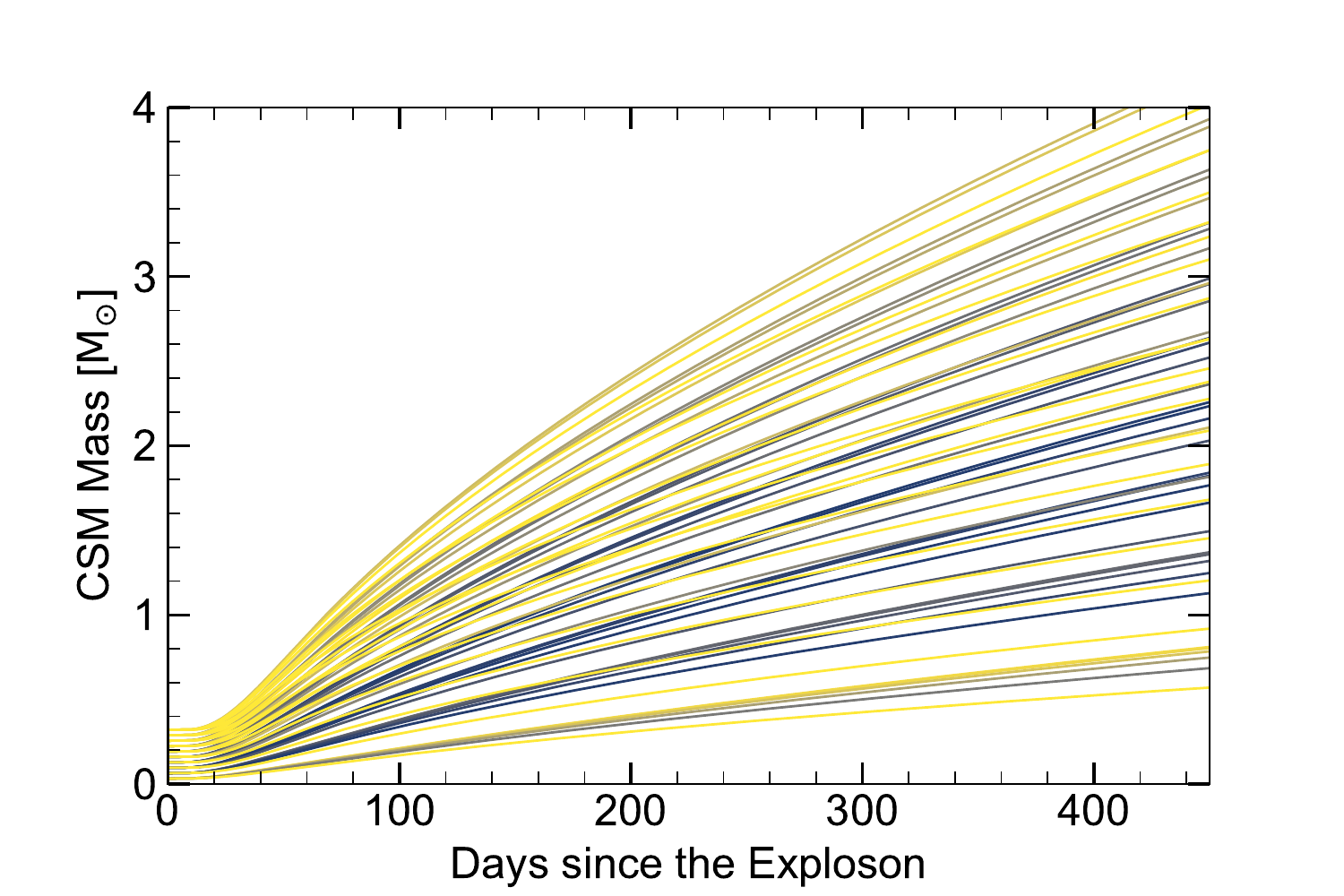}{0.45\textwidth}{(d)}
          }
\caption{The same as Figure \ref{fig:6}, but for $R_{\rm CSM, in} = 3.0\times 10^{15}{\rm cm}$.}
\label{fig:app3}
\end{figure*}

Figures \ref{fig:app2} and \ref{fig:app3} show the results of the light curve modeling for other parameters;  $R_{\rm CSM, in} = 1.0\times 10^{15}{\rm cm}$ and $3.0\times 10^{15}{\rm cm}$.

\bibliography{manuscript}{}
\bibliographystyle{aasjournal}

\end{document}